\documentclass[twocolumn]{aastex63}
\usepackage{bm}
\usepackage{multirow}
\usepackage{color}
\usepackage{float}
\usepackage{tabularx}
\makeatother
\makeatletter
\usepackage{comment}
\usepackage{natbib}
\usepackage{amsmath}
\usepackage{amsfonts}
\usepackage{amssymb}
\usepackage{epsfig}
\usepackage{graphicx}
\usepackage{booktabs}
\usepackage{array}
\usepackage{hyperref}
\hypersetup{colorlinks=true, citecolor=blue, filecolor=magenta,
 urlcolor=cyan, linkcolor=blue}

\shortauthors{Long et al.}

\submitjournal{ApJ, first round of revisions incorporated}

\begin{document} 

\title{Emergence of an Ultra-Red Ultra-Massive Galaxy Cluster Core at \lowercase{$z=4$}}

\author{Arianna S. Long}
\affiliation{Department of Physics and Astronomy, University of California, Irvine, CA 92697, USA}
\email{arianna.long@uci.e du}
\author{Asantha Cooray}
\affiliation{Department of Physics and Astronomy, University of California, Irvine, CA 92697, USA}
\author{Jingzhe Ma}
\affiliation{Department of Physics and Astronomy, University of California, Irvine, CA 92697, USA}
\author{Caitlin M. Casey}
\affiliation{Department of Astronomy, The University of Texas at Austin, 2515 Speedway Boulevard Stop C1400, Austin, TX 78712, USA}
\author{Julie L Wardlow}
\affiliation{Physics Department, Lancaster University, Bailrigg, Lancaster, LA1 44B, UK}
\author{Hooshang Nayyeri}
\affiliation{Department of Physics and Astronomy, University of California, Irvine, CA 92697, USA}
\author{R. J. Ivison}
\affiliation{European Southern Observatory, Karl-Schwarzschild-Strasse 2, D-85748 Garching, Germany}
\author{Duncan Farrah}
\affiliation{Department of Physics and Astronomy, University of Hawaii, 2505 Correa Road, Honolulu, HI 96822, USA}
\affiliation{Institute for Astronomy, 2680 Woodlawn Drive, University of Hawaii, Honolulu, HI 96822, USA}
\author{Helmut Dannerbauer}
\affiliation{Instituto de Astrof\'isica de Canarias, E-38205 La Laguna, Tenerife, Spain}
\affiliation{Departmento de Astrof\'isica, Universidad de La Laguna, E-38206 La Laguna, Tenerife, Spain}

\begin{abstract}
Recent simulations and observations of massive galaxy cluster evolution predict that the majority of stellar mass build up happens within cluster members by $z=2$, before cluster virialization. Protoclusters rich with dusty, star-forming galaxies (DSFGs) at $z>3$ are the favored candidate progenitors for these massive galaxy clusters at $z\sim0$. We present here the first study analyzing stellar emission along with cold dust and gas continuum emission in a spectroscopically confirmed $z=4.002$ protocluster core rich with DSFGs, the Distant Red Core (DRC). We combine new \textit{HST} and \textit{Spitzer} data with existing Gemini, \textit{Herschel}, and ALMA observations to derive individual galaxy-level properties, and compare them to coeval field and other protocluster galaxies. All of the protocluster members are massive ($>10^{10}$\,M$_\odot$), but not significantly more so than their coeval field counterparts. Within uncertainty, all are nearly indistinguishable from galaxies on the star-forming vs. stellar mass main-sequence relationship, and on the star formation efficiency plane. Assuming no future major influx of fresh gas, we estimate that these gaseous DSFGs will deplete their gas reservoirs in $\sim300$\,Myr, becoming the massive quiescent ellipticals dominating cluster cores by $z\sim3$. Using various methodologies, we derive a total $z=4$ halo mass of $\sim10^{14}$\,M$_\odot$, and estimate that the DRC will evolve to become an ultra-massive cluster core of mass $\gtrsim10^{15}$\,M$_\odot$ by $z=0$. \\
\end{abstract}


\section{Introduction}
Environmental impacts on galaxy evolution are best understood at $z<2$.  In observational studies on the local Universe all the way out to $z=1.5$, galaxy clusters are known to host excess populations of red and massive galaxies when compared to coeval field counterparts \citep{lewis02,wake05,patel09,vander13,scoville13,lem19}. In order to form these massive, quiescent populations, studies suggest that clusters must form the majority of their mass ($\sim50\%$) and initiate rapid quenching by $z=2$ \citep[e.g.][]{cooper10, papovich10, rettura11}, which means clusters at $z\gtrsim3$ should host many actively star-forming galaxies \citep{contini16,chiang17}. Unfortunately, observational selection biases bear inconclusive results on whether there exists an excess of star-formation activity in early cluster environments at $z>2$ \citep[e.g.][]{steidel05,koyama13,cooke16}. This is likely due to the fact that the methods originally developed to find clusters were inherently built to detect near-virialized clusters at $z\lesssim2$ with strong red sequences already in place \citep[e.g.][]{gladders00,rosati2002,eisen08,saro09,chiang13} and/or evidence of a hot X-ray emitting intracluster medium \citep[e.g.][]{rosati2002, mullis05, willis13, bleem15}.

Full characterization of $z>3$ early cluster, aka protocluster \citep{overzier16}, environments is vital to our efforts in understanding several cosmological processes, including the collapse of filamentary structures \citep[e.g.][]{ssa22}, the formation and assembly of massive halos in $\mathrm{\Lambda CDM}$ \citep[e.g.][]{suwa06,hc12}, and the births of the most massive galaxies in the Universe: brightest cluster galaxies \citep[e.g.][]{del07,rf18,cooke19,rennenhan}. Currently, most protoclusters cataloged at $z>3$ are discovered and characterised based on their rest-frame optical/UV emission owing to their selection techniques (e.g. systematic narrow-band/spectroscopic searches for overdensities of Ly$\alpha$ emitters, H$\alpha$ emitters, and/or LBGs, e.g. \citet[][]{ven07,daddi09,capak11,koyama13,lem18,hig19}). However, these techniques are blind to a rare but important phase of massive galaxy evolution that contributes immensely to cosmic star formation: dusty, star-forming galaxies (DSFGs, see \citet*{casey2014dusty} for a review; see also HAE229 in \citet{doh10,dann14}, and \citet{dann17}).

Recent far-IR and sub-millimeter observations have uncovered populations of dusty, star-forming galaxies residing in overdense environments at $z>2$ \citep[e.g.][]{geach06, chapman09, dann14, clements14, casey15, um15, hung16, miller18, gg19, harikane19, lac19}. Their incredible bursts of star formation over short periods of time at high-$z$ makes DSFGs ideal candidates for driving rapid stellar mass build up at $z>3$ in protoclusters, before the widespread onset of a red sequence is in place. The strong presence of DSFGs in these overdensities is not a coincidence, but likely a key part of protocluster evolution \citep{casey16}.

Detailed multiwavelength characterization has been carried out at the individual galaxy level for many $z\lesssim3$ nearly-virialized protoclusters with DSFGs \citep[e.g.][]{spitler12,casey15,um15, dann17}, but most newly discovered $z\gtrsim3$ protoclusters with DSFGs either (a) have only a handful (1-3) of these rare starbursts \citep{daddi09,capak11,walter12,pavesi18}, (b) are DSFG-rich but with resolved observations limited to only their far-IR and sub-mm properties \citep{miller18, hill2020}, or (c) are not yet spectroscopically confirmed members of the protocluster \citep{champ18,harikane19}. In order to fully assess the role and evolution of DSFGs in overdense environments (e.g. are they the primary progenitors of BCGs or other massive spheroidals seen in modern day clusters?), we must seek and then explore the properties of these rare and extreme environments across the energy spectrum. 

The work presented here is the first to link resolved stellar emission with cold dust and gas from star-forming regions in a spectroscopically confirmed $z=4$ protocluster rich with DSFGs. We combine high-resolution \textit{Hubble Space Telescope} (\textit{HST}) data, deep Gemini FLAMINGOS-2 data, and deep \textit{Spitzer} IRAC observations to probe the rest-frame UV of an extremely dense protocluster core spectroscopically confirmed at $z=4.002$ with the \textit{Atacama Large Millimeter/sub-millimeter Array} (ALMA): the Distant Red Core (DRC). The DRC was identified by \citet{ivison2016} as the single reddest source in a systematic search for high-$z$, extreme star-forming systems in the $\approx600$\,deg$^2$ \textit{Herschel} Astrophysical Terahertz Large Area Survey \citep[\textit{H}-ATLAS,][]{eales} based on ``red'' {\it Herschel} SPIRE flux densities ($S_{500} > S_{350} > S_{250}$, \citet{ivison2016,asboth16}). Follow up APEX LABOCA 870\,$\mu$m imaging across a 10$^\prime$ field confirmed a significant ($2.15^{+0.8}_{-0.5}$) overdensity of DSFGs \citep{lewis18}, and subsequent ALMA 2-3\,mm spectroscopic scans on the two brightest 870\,$\mu$m emitters resolved an astounding 10 DSFGs at $z_{spec}=4.002$ within a 260\,kpc$\times$310\,kpc$\times$87\,Mpc region (\citet{oteo18}, see also \citet{fudamoto17}), making the DRC core one of the rarest and most dense concentrations of DSFGs known at high-$z$ (see also \citet{miller18} for a similar structure at $z=4.3$).

In Section \ref{sec:data}, we present multiwavelength data and counterpart identification; in Section \ref{sec:SEDs}, we describe the spectral energy distribution (SED) fitting process; in Section \ref{sec:indv} we present our results comparing individual protocluster members to field galaxies, and in Section \ref{sec:dmh} we discuss the DRC in context of global galaxy cluster evolution; we summarize our conclusions in Section \ref{sec:concl}. Throughout this paper, we adopt a cosmology of H$_0=70$\,km\,s$^{-1}$\,Mpc$^{-1}$, $\Omega_M=0.3$, and $\Omega_\Lambda=0.7$.

\begin{figure*}
\includegraphics[trim=0cm 0cm 0cm 0cm, scale=0.55, width=1\textwidth]{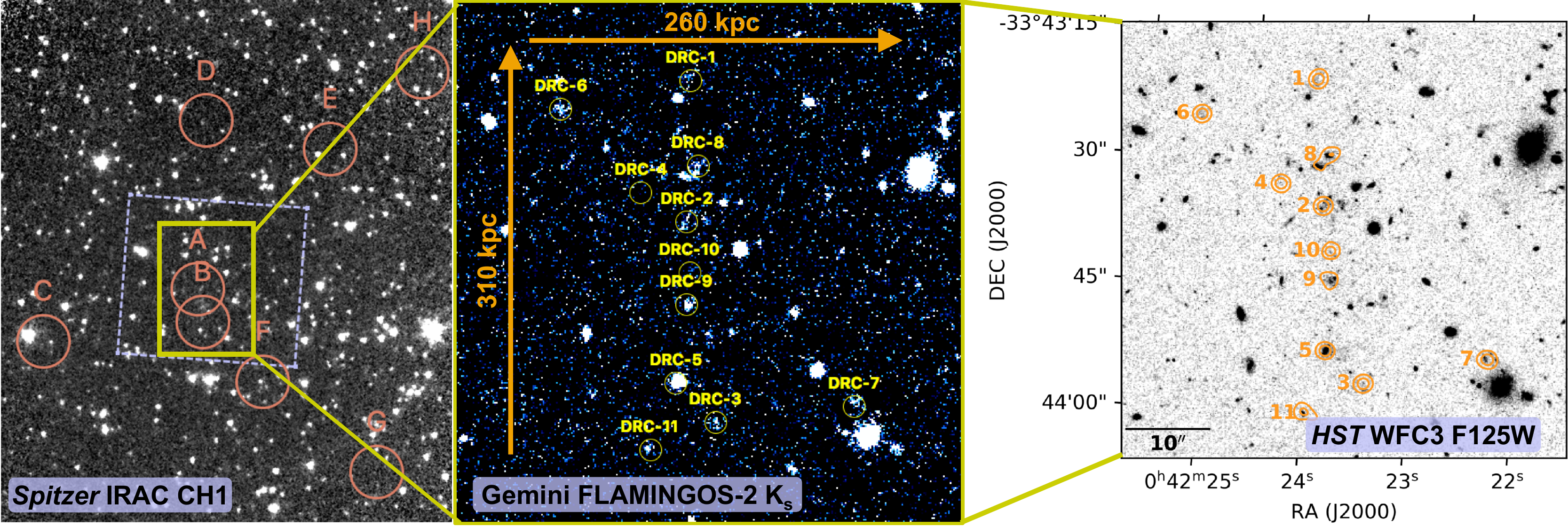}
\includegraphics[trim=0.1cm 0.1cm 0.1cm 0cm, width=1\textwidth]{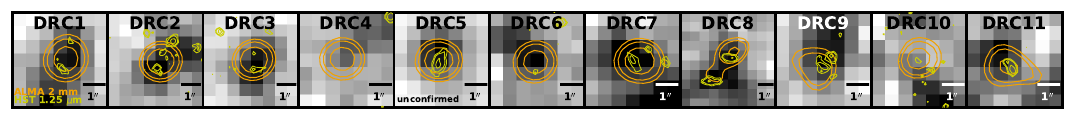}
\caption{\textit{Top}: On the left is a zoomed out \textit{Spitzer} IRAC 3.6\,$\mu$m image with the positions of other potential protocluster members outside of the DRC (objects C-H, as seen in 870\,$\mu$m LABOCA imaging in \citet{oteo18}) circled in pink. The blue dashed box shows the \textit{HST} 1.25\,$\mu$m image footprint over the core of the protocluster (the DRC). The middle panel is a zoomed in Gemini FLAMINGOS-2 K$_s$-band image of the protocluster core, with ALMA positions for each DRC component encircled in green. Finally, on the right we present a zoomed in image of the DRC as seen by \textit{HST} at 1.25\,$\mu$m. Overlaid in orange are ALMA 2\,mm contours at 2$\sigma$ and 5$\sigma$. DRC-4 is attenuated in the Gemini and \textit{HST} images, as is DRC-10. \textit{Bottom}: Observed-frame \textit{HST} 1.25\,$\mu$m and ALMA 2\,mm continuum contours overlaid on \textit{Spitzer} IRAC $3.6\,\mu$m images for all DRC components (regardless of positive near-IR detection). ALMA contour levels (orange) are at 2, 3, and 5$\sigma$; \textit{HST} contour levels (yellow) are at 2.5, 3.5, and 7$\sigma$. At $z=4.002$, an arcsecond corresponds to a spatial scale of $\sim$7.1 kpc. In this study, we consider objects within $1.14^{\prime\prime}$ ($\sim8$\,kpc) of the ALMA centroid to be the collective rest-frame optical/near-IR counterpart. For nearly all objects, this includes only the 1.25\,$\mu$m bright objects within the shown ALMA contours, except for DRC-8 where we include the additional near-IR bright galaxy in the southwest region due to the overlapping shape of the ALMA contours. See section \ref{identifying} for more details.
\footnotesize 
}
\label{fig:insetcontours}
\end{figure*}

\section{Observations} \label{sec:data}

\subsection{HST WFC3}
In \textit{HST} Cycle 25, we used the F125W filter to observe a subset of eight ultra-red \textit{Herschel} objects with precise coordinates from ALMA observations and clear \textit{Spitzer} IRAC counterparts (PID: 15464, PI: A. Brown). We used a four-point dither pattern with a 653\,s exposure per frame, achieving a total on-source integration time of 43.5 minutes over the F125W band. We use the final calibrated data from the Barbara A. Mikulski Archive for Space Telescopes (MAST), which is combined and corrected using the standard WFC3 reduction pipeline (\texttt{calwf3} v3.4.2 and \texttt{DrizzlePac} v2.1.21). At this wavelength (1.25\,$\mu$m), we determine 3$\sigma$ depths of m$_\mathrm{AB}=22.5$ on point-like sources and a \textsc{psf} of $0.18^{\prime\prime}$.

\subsection{Gemini FLAMINGOS-2}
In 2014, the FLAMINGOS-2 instrument on the Gemini-South telescope observed the DRC for a total of $\sim$4 hours in the $K_s$-band (PID: GS-2014A-Q-58, PI: L. Dunne). Here, we use the same reduced data presented in \citet{oteo18} (Section 2.5, therein), which reaches a final 3$\sigma$ depth of m$_\mathrm{AB}=25$ with an average seeing of $0.72^{\prime\prime}$.

\subsection{Spitzer IRAC}
In Cycle 13, as part of a follow-up campaign to measure the rest-frame optical emission for 300 $z\gtrsim4$ ultra-red DSFGs, the \textit{Spitzer} Space Telescope Infrared Array Camera (IRAC) imaged the DRC at 3.6\,$\mu$m and 4.5\,$\mu$m (PID: 13042, PI: A. Cooray, see also \citet{maspitzer}). Images in each band were taken over a 36-point dither pattern with a 30\,s exposure per frame, achieving a total integration time of 18 minutes per band. We use the reduced post-basic calibrated data (pBCDs) from the \textit{Spitzer} Science Center (vS19.2), achieving depths of m$_\mathrm{AB}=24$ and 26, and \textsc{pst} limits of $0.93^{\prime\prime}$ and $1.13^{\prime\prime}$, respectively.

\subsection{Herschel SPIRE}
Data Release 2 of the \textit{H}-ATLAS survey \citep{eales,valiante,maddox} captured the DRC at rest-frame far-IR wavelengths. SPIRE observations were taken in parallel over the South Galactic Pole, with \textsc{fwhm}s of 17.8$^{\prime\prime}$, 24.0$^{\prime\prime}$, and 35.2$^{\prime\prime}$ at 250, 350, and 500\,$\mu$m, respectively. Ultra-red sources were selected with a 3.5$\,\sigma$ detection threshold at 500\,$\mu$m flux densities (S$_{500}$) $ > 30$\,mJy, with S$_{500}$/S$_{250} \ge1.5$ and S$_{500}$/S$_{350} \ge0.85$ \citep{ivison2016}. We refer the reader to \citet{ivison2016}, \citet{valiante}, and \citet{maddox} for extensive details on observations and source extraction, and to Section \ref{sec:spire} for details on deblending \textit{Herschel} SPIRE data.

\subsection{ALMA}
As presented in \citet{oteo18}, successful spectroscopic confirmation of DRC members required several spectral scans using the \textit{Atacama Large Millimeter Array} (ALMA) to unambiguously detect more than one emission line. We refer the reader to \citet{oteo18} for the full chronicle, and briefly summarize the data used in this work below.

The DRC core has a spectroscopic redshift of $z_\mathrm{spec}=4.002$, determined via ALMA 2\,mm spectral scans (PID: 2016.1.01287.S, PI: I. Oteo) carried out over two pointings, with an average synthesized beam size of $1.6^{\prime\prime}$. All sources but DRC-5 were spectroscopically confirmed via detection of $^{12}$CO(6--5) emission, and up to four additional emission lines detected for some of these objects (including [C\,\textsc{i}](1--0), H$_2$O($2_{11}-2_{02}$), $^{12}$CO(4--3), and $^{12}$CO(2--1); PID: 2013.1.00449.S, P.I. A. Conley; PID: 2013.A.00014.S, PI: R.J. Ivison; and PID: 2013.1.00449.S, PI: R.J. Ivison). At $z=4.002$, the respective field of view for the 2\,mm mosaic is roughly $675\,\mathrm{kpc}\,\times\,433$\,kpc with a physical synthesized beam size of 11.4\,kpc; thus, these sources are unresolved at sub-mm wavelengths (with the exception of DRC-1 which was imaged with 0.12$^{\prime\prime}$ resolutions at 870$\,\mu$m in PID: 2013.1.00001.S, PI: Ivison; this data is not included in this analysis).

\section{Photometry and Counterpart Selection}
Here we review the photometry and counterpart selections used in this analysis. In Section \ref{identifying}, we discuss how we carry out near-IR counterpart identification for each DRC component, and in Sections \ref{sec:tphot} and \ref{sec:spire}, we discuss the deblending techniques used to derive fluxes for each DRC component in the \textit{Spitzer} IRAC data \textit{Herschel} SPIRE data, respectively. The resulting photometry is tabulated in Table \ref{table:observations}.

\subsection{Near-Infrared Photometry} \label{sec:phot}
For all observed-frame near-IR data (\textit{HST}, Gemini, and \textit{Spitzer}), we use the \textsc{source extractor} package \citep{setractor} in single-image mode to identify objects, and assign total fluxes based on \texttt{FLUX\_ISO} values, as many sources had disturbed morphologies not easily identified by elliptical projections. We compare our photometry for several stars also in the 2MASS catalogs \citep{2mass} and find a $<10\%$ difference in flux density estimates. In the following paragraph, we discuss the signal-to-noise limits employed throughout this work.

For \textit{HST} data, we measure 9/11 DRC sources at S/N$\,\gtrsim2$. In this band, only 5/11 sources have S/N$\,>3$; we choose to keep the additional four objects with $3>$\,S/N$\,\gtrsim2$ as these objects are clear (S/N$\,=4-5$) detections at in the K$_s$ band (2.2\,$\mu$m), and the positional offsets between the sources as seen in the F125W and K$_s$ filters are $\le0.4^{\prime\prime}$ (which is $\sim$1-2 pixels or less in the Gemini image). For the deeper FLAMINGOS-2 data, we detect 9/11 DRC objects at S/N$\,>3$, with the remaining two objects, DRC-4 and 10, at S/N$\,<2$ (which is consistent with HST). In the 3.6\,$\mu$m and 4.5\,$\mu$m \textit{Spitzer} images, 8/10 and 9/10 DRC components are detected at S/N$\,>3$, respectively. However, 6/10 of these objects are blended with neighboring sources. In Section \ref{sec:tphot}, we describe the process for deblending the IRAC counterparts with their neighbors. 

\subsection{Identifying HST and Gemini Counterparts} \label{identifying}
Upon inspection, many DRC members break apart into several rest-frame UV ($\lambda = 2500\,\mathrm{\AA}$) counterparts within their respective ALMA contours, several of which exhibit clumpy and/or interacting morphologies (Figure \ref{fig:insetcontours}). These morphologies are expected for the majority of galaxies at high-redshifts ($z>1$), due to increased merger fractions and star formation activity during this epoch of the Universe \citep[e.g.][but see also \citet{hodge16}]{cowie95,vanden96,elm04,elm07,agertz09,dekel06,bournard14,shibuya16}. Moreover, studies suggest that the bright sub-mm flux from DSFGs hails not just from isolated starbursts, but also from merger-induced starbursts and/or pairs of galaxies (not necessarily individually bursting) undergoing a spiral infall \citep{guowhite,dave2010,hopkins2010,gonzalez2011,hayward2011,hayward2012,hayward2013b,somerville2012,narayanan15,chen15, gg18,cowie18,mcalpine} -- and, in overdense regions like that of protocluster environments, there is an increased merger fraction compared to coeval field environments \citep{klypin2001,fakhouri2009,lotz2013,hine2016}. 

Considering the aforementioned evidence, and the large ALMA beam sizes relative to the HST resolution, we decide to treat each ALMA object as it's own global physical star-forming system, capturing all observed-frame near-IR bright objects within a physically motivated radius on the order of galactic scales: $\sim8$\,kpc ($1.14^{\prime\prime}$ at $z=4$, also seen in \citet{wiklind14}). This chosen radius emits from the center of the ALMA 2\,mm emission for each object, within which we deem all rest-frame UV bright objects as a cumulative counterpart. We note that, for many of the sources with multiple near-IR counterparts, the center of the ALMA emission does not align with a singular near-IR bright object. Instead, it is often centered between two or more objects, which is unsurprising considering that dust and stellar offsets are not uncommon in DSFGs \citep[e.g.][]{chen15, casey17}. The physical distance chosen ensures we capture only closely interacting pairs, individual galaxies dominated by patches of star-forming regions / giant molecular clouds, and/or systems with irregularly shaped dust and stellar offsets due to recent gravitational interactions or to strong dust extinction \citep[seen in e.g.][]{wiklind14,chen15,casey17,gg18,cowie18}. The only exception to this case is DRC-8 in which we choose to include the additional rest-frame UV object $\sim2^{\prime\prime}$ to the southwest of the brightest part of the ALMA centroid as the ALMA observations appear to also detect this additional object. 

For 5/11 sources (nos 1, 2, 7, 8, and 9), more than one \textit{HST} counterpart is found within the 1.14$^{\prime\prime}$ (8\,kpc) radius (see Figure \ref{fig:insetcontours}). With the available data, we cannot definitively rule out the possibility of low-redshift interlopers in the optical/near-IR data. However, for DRC objects 1,2,7, and 8, which show multiple possibly interacting components, the \textsc{fwhm} of the CO($6-5$) emission used to originally spectroscopically confirm cluster membership is extremely broad ($>1000$\,km\,s$^{-1}$, see Figure 2 in \citet{oteo18}). We interpret this as evidence that these four objects are likely ongoing merger events, and therefore the 2-3\,mm continuum measurements represent star-formation triggered within the global system. Object 9 does not have the broad emission, but shows morphologies indicative of a disturbed system with possible dust offsets from the preceding interactions. 

For the 5/11 sources with more than one \textit{HST} counterpart,we sum the respective fluxes to form a total observed-frame 1.25\,$\mu$m flux for each ALMA DRC component -- still only including \textit{HST} sources with S/N\,$\gtrsim2$. Uncertainties from multiple counterparts are added in quadrature. The Gemini observation, affected by seeing, is more blended than the \textit{HST} image. So, where necessary, we repeat this exact method for multiple objects detected within the same radius in the K$_s$-band image, although this only applies to two sources: DRC-2 and DRC-7. As mentioned earlier in this section, DRC-8 also includes two objects in the \textit{HST} and Gemini flux density measurements (both at S/N\,$>3$), with the uncertainties added in quadrature.

We note that these assumptions could result in an overestimation of the stellar component in the SED fitting process, and thus we interpret the resulting properties as loose estimates and take care to include all uncertainties in our analyses and figures throughout this work.

\begin{deluxetable*}{lcccccccccccccc}
    \centering
    \tablecaption{Measured flux densities for each cluster member. Measurements listed without uncertainties are used as upper limits. See Section \ref{sec:phot} and \ref{sec:spire} for more details.}
    \tablehead{
    \colhead{ID} & \colhead{S$\mathrm{_{1.25\,\mu m}}$ } & \colhead{S$\mathrm{_{2.2\,\mu m}}$} & \colhead{S$\mathrm{_{3.6\,\mu m}}$} & \colhead{S$\mathrm{_{4.5\,\mu m}}$} & \colhead{S$\mathrm{_{250\,\mu m}}$} & \colhead{S$\mathrm{_{350\,\mu m}}$} & \colhead{S$\mathrm{_{500\,\mu m}}$} & \colhead{S$\mathrm{_{2\,mm}}$} & \colhead{S$\mathrm{_{3\,mm}}$}\\
     & \colhead{[$\mu$Jy]} & \colhead{[$\mu$Jy]} & \colhead{[$\mu$Jy]} & \colhead{[$\mu$Jy]} & \colhead{[mJy]} & \colhead{[mJy]} & \colhead{[mJy]} & \colhead{[$\mu$Jy]} & \colhead{[$\mu$Jy]} \\
     }
    \startdata
    DRC-1 & 2.46\,$\pm\,1.03$ & 4.16\,$\pm\,0.76$ & 6.11\,$\pm\,1.28$ & 5.73\,$\pm\,0.97$ & 11.14\,$\pm\,5.54$ & 34.34\,$\pm\,7.59$ & 45.23\,$\pm\,9.24$ & 2117\,$\pm\,58$ & 406\,$\pm\,28$ \\
    DRC-2 & 8.56\,$\pm\,2.26$ & 4.02\,$\pm\,0.86$ & 11.21\,$\pm\,3.69$ & 4.89\,$\pm\,1.46$ & 16.22\,$\pm\,7.53$ & $<15.64$ & $<10.35$ & 723\,$\pm\,11$ & 154\,$\pm\,10$ \\   
    DRC-3 & 1.39\,$\pm\,0.77$ & 5.14\,$\pm\,0.88$ & 5.49\,$\pm\,1.64$ & 6.60\,$\pm\,1.58$ & 12.43\,$\pm\,5.94$ & 24.85\,$\pm\,6.42$ & 18.38\,$\pm\,9.11$ & 659\,$\pm\,10$ & 218\,$\pm\,22$ \\
    DRC-4 & $<1.0 $ & $<0.40$  & $<2.10 $ & $<1.29 $ & $<7.18$  & $<10.28$  & $<7.18$  & 347\,$\pm\,99$ & 75\,$\pm\,17$\\
    DRC-5\tablenotemark{a} & 12.98\,$\pm\,2.31$ & 18.41\,$\pm\,0.73$ & 19.12\,$\pm\,3.44$ & 20.86\,$\pm\,3.54$ & $<3.23$ & $<6.67$ & $<3.78$ & 295\,$\pm\,94$ & 110\,$\pm\,12$\\
    DRC-6 & 1.64\,$\pm\,0.83$ & 4.62\,$\pm\,1.16$ & 4.62\,$\pm\,0.74$ & 3.80\,$\pm\,0.53$ & $<6.59$ & $<4.55$ & $<4.72$ & 282\,$\pm\,65$ & 102\,$\pm\,11$ \\
    DRC-7 & 7.32\,$\pm\,1.76$ & 4.34\,$\pm\,0.58$ & 6.67\,$\pm\,1.60$ & 6.86\,$\pm\,1.37$ & $<3.63$ & $<6.25$ & $<8.71$ & 176\,$\pm\,82$ & --\\
    DRC-8 & 16.89\,$\pm\,3.71$\tablenotemark{b} & 3.81\,$\pm\,0.70$ & 5.09\,$\pm\,0.99$ & 3.20\,$\pm\,1.06$ & $<4.34$ & $<6.79$ & $<7.74$ & 55\,$\pm\,10$ & -- \\
    DRC-9 & 11.67\,$\pm\,2.22$ & 3.85\,$\pm\,0.58$ & 4.27\,$\pm\,1.28$ & 3.74\,$\pm\,0.93$ & $<5.22 $ & $<5.50$ & $<3.21$ & 42\,$\pm\,11$ & -- \\
    DRC-10 & $<1.16$  & $<0.59$  & 2.78\,$\pm\,0.42$  & 1.50\,$\pm\,0.30$ & $<5.98 $ & $<5.09$ & $<3.20$ & 40\,$\pm\,7$ & -- \\
    DRC-11 & 3.47\,$\pm\,1.19$ & 2.82\,$\pm\,0.56$ & 3.73\,$\pm\,0.56$ & 2.55\,$\pm\,0.36$ & $<2.98$ & $<7.21$ & $<4.60$ & 39\,$\pm\,9$ & -- \\
    \enddata
    \label{table:observations}
    \tablenotetext{a}{DRC-5 is not spectroscopically confirmed at $z=4.002$ like the other 10 members. We still include this object in our analyses in Sections \ref{sec:indv} and \ref{sec:dmh}, and note any impacts on the global cluster properties if DRC-5 is indeed not a true member of this cluster core.}
    \tablenotetext{b}{Total combined flux for the two galaxies lying within the ALMA contours seen in the \textit{HST} image; at 2.2\,$\mu$m, the two components were recognized as one singular object.}
\end{deluxetable*}

\subsection{Generating Respective Herschel Flux Densities} \label{sec:spire}

In each of the \textit{Herschel} SPIRE images, the DRC is blended together as a single object. The protocluster was systematically selected as an ``ultra-red'' source based on it's rising SPIRE flux densities ($S_{500} > S_{350} > S_{250}$, \citet{ivison2016}) believed to trace the Wien side of the far-infrared blackbody for galaxies at $z\gtrsim4$. \citet{ivison2016} measured a total flux for the DRC (aka SGP-354388) of $26.6\pm8.0$, $39.8\pm8.9$, and $53.5\pm9.8$ mJy at 250, 350, and 500\,$\mu$m, respectively; additional follow-up SCUBA-2 and LABOCA 850-870\,$\mu$m measurements also resolved the DRC (but this data is not used in this analysis). In the following, we describe how we determine individual object flux densities or upper limits.

Using ALMA positional priors for each DRC component, we deblend the far-IR emission with the probabilistic deblender XID+ \citep{xid}.\footnote{http://herschel.sussex.ac.uk/XID\_plus} XID+ is a tool designed specifically to deblend SPIRE maps using higher-resolution positional priors and a Bayesian inference to obtain the full posterior probability distribution function on flux estimates.

When all 11 sub-mm bright objects are considered in the fit, the results produce flux densities $<10$\,mJy with S/N\,$\sim1$ for each source. These estimates are considerably close to (or below) the reliability thresholds defined in \citet{xid} (5 and 10\,mJy for 250 and 350-500\,$\mu$m, respectively) and might indicate that none of these galaxies would be detected individually in \textit{Herschel} surveys if they were separated. However, XID+ is reliant on the high-res positional priors of known dusty objects, which we have with ALMA data, and we also know that all objects (except DRC-5) sit at the same redshift; this means that objects that are brightest in the 2\,mm observations are likely more massive/dust-rich than their fainter co-cluster members. Thus, we performed another deblending fit using only the six objects brightest at 2\,mm - this produced similar results as the 11-object fit. Finally, when iterating XID+ on the \textit{four} brightest 2\,mm objects only (nos. 1-4), we recover the majority of the SPIRE flux with estimates above the reliability threshold and at S/N\,$\gtrsim2$ significance for members 1, 2, and 3. We interpret this ensemble of fits to mean that objects 1-3 are likely the main contributors to the \textit{Herschel} SPIRE fluxes, and that contributions from the other sources are negligible/undetectable at the shallow depths of this survey. This is also found to be true in the \citet{smith19} ALMA detected cluster, where the majority of the \textit{Herschel} and SCUBA-2 sub-mm flux was be attributed to the three (out of ten) brightest ALMA sources. 

We adopt the fluxes as detections in the SED fitting process for objects 1-3, and use the results from the first pass (that included all 11 sources) as generous upper limits in the SED-fitting process for the remaining individual galaxies. In the \hyperref[sec:appendix]{Appendix}, we show the best-fit SEDs and discuss the galaxy properties for objects 1-3 that result from a fit using the (smaller) upper limits derived in the first XID+ pass (which used all 11 sources instead of four). In general, when using the upper limits for all sources, we find no major differences on the implications discussed in this analysis for these galaxies or for the cluster as a whole.

We note that without these SPIRE upper limits / flux density estimates, our SED models generate a much larger far-IR component for each member (with flux densities on the order of $5-10\times$ larger than what the deblended values predict). Furthermore, we find that SED fits using the deblended values produce galaxy-level properties that are within 1$\sigma$ uncertainty of those found in \citet{oteo18} (which were generated by fitting the 2\,mm data to ALESS SED templates). Without them, the resulting stellar masses and dust luminosities are much greater (2-10$\times$ greater, which is unphysical in several cases). Thus, we find these upper limits and fluxes critical to our SED-fitting process.

\begin{figure}
\includegraphics[trim=0cm 0cm 0cm 0cm, scale=1.]{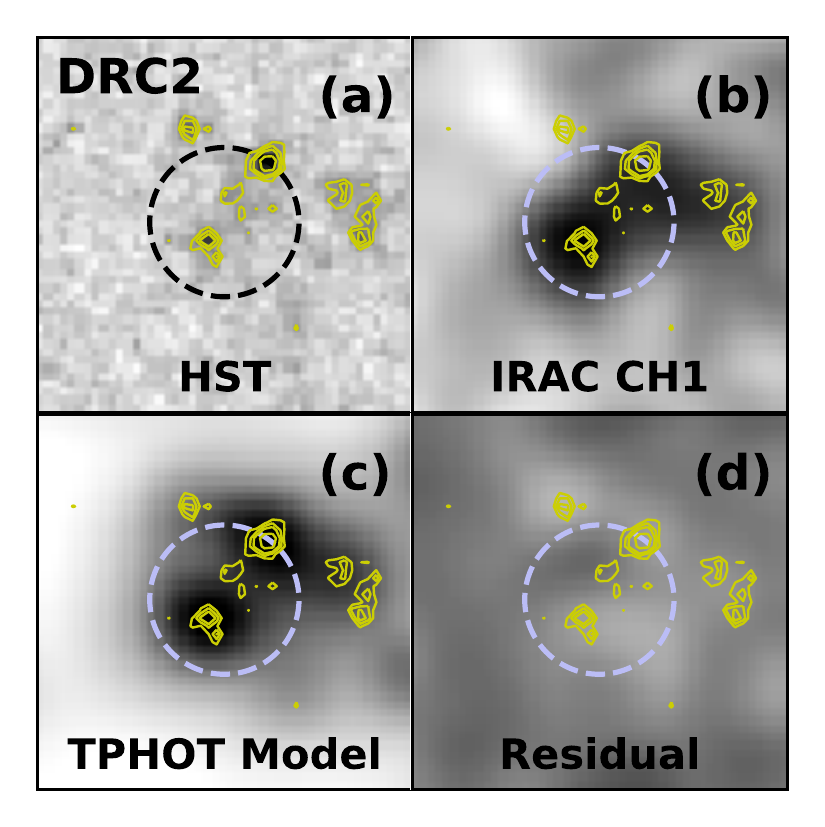}
\caption{A progression example of the \textsc{tphot} modeling and deblending process for DRC-2. Contours match the \textit{HST} contour stamps in Figure \ref{fig:insetcontours}. The dashed circle represents a radius of 1.14$^{\prime\prime}$ in which all sources within are considered a cumulative near-IR counterpart to the ALMA centroid; this includes the source sitting on the circle for DRC-2. We use \textsc{tphot} to deblend IRAC fluxes for protocluster members that appear blended with nearby neighbors outside this radius (see Section \ref{sec:tphot} for more details). (a) is the \textit{HST} image of the disturbed DRC-2; (b) is the corresponding \textit{Spitzer} IRAC 3.6\,$\mu$m image that's been \textsc{swarp}-ed \citep{swarp} to match the \textit{HST} pixel resolution; (c) is the modeled IRAC image \textsc{tphot} creates using \textit{HST} positional priors and a PSF convolution kernel, and (d) is the residual flux remaining after extraction of the modeled flux from the original IRAC map. We see no systemic issues in our residual maps and recover $85-100\%$ of the original flux for sources with clear singular IRAC counterparts (e.g. members 5 and 7), and thus consider our deblended fluxes reliable. See Section \ref{sec:tphot} for more details.
\footnotesize 
}
\label{fig:tphot}
\end{figure}

\begin{figure*}
\begin{center}
\includegraphics[trim=0.3cm 0.4cm 0cm 0cm, scale=2.75]{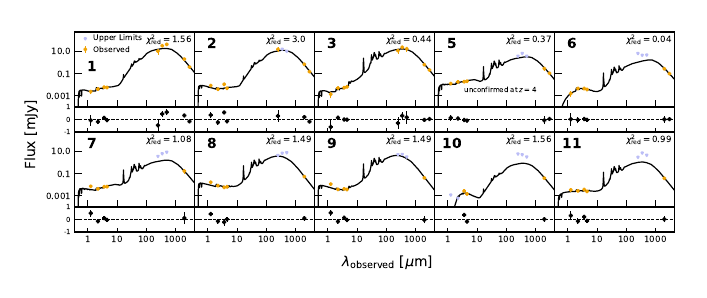}
\end{center}
\caption{Best fit spectral energy distributions from \textsc{cigale} for each member considered in this paper. Observed fluxes (including those that are found via deblending) are plotted as orange dots, while upper limits are plotted as blue triangles. We use the combined flux densities for all \textit{HST} and Gemini sources within 1.14$^{\prime\prime}$ of each of the DRC's ALMA centroids (see Section \ref{identifying}). In the top left is the ID for each member and in the top right is the reduced $\chi^{2}$ value for each fit. Beneath each SED is the relative percent difference between the observed and model fluxes. For several sources, DRC-2 in particular, there is a visible offset between the modeled flux and the 3.6\,$\mu$m measured flux; this emission is measured in a band wide enough to capture redshifted H$\alpha$ flux, which is expected in excess in a highly disturbed, bursty system such as this. DRC-4 is not shown as only upper limits were measured in all bands. See Section \ref{sec:SEDs} for SED fitting details.
\footnotesize 
}
\label{fig:seds}
\end{figure*}

\subsection{Deblending IRAC Counterparts} \label{sec:tphot}
For several objects (e.g. DRC-2, DRC-8, and DRC-9), the 3.6$\,\mu$m and 4.5$\,\mu$m flux is blended with nearby sources outside of the projected $1.14^{\prime\prime}$ merging radius. To avoid overestimating stellar properties, we deblend the IRAC photometry using \textsc{tphot} \citep{tphot1,tphot2}, a software package designed to extract and deblend photometry from low resolution images (IRAC) using high resolution priors (HST). We use \textsc{Source Extractor} \citep{setractor} to generate the relevant input catalogs and segmentation maps, then we apply \textsc{swarp} \citep{swarp} to both IRAC images to match the HST/F125W image pixel resolution (0.127$^{\prime\prime}$). Finally, we use \textsc{pypher} to generate a convolution kernel between the IRAC and \textit{HST} PSFs. To achieve optimal performance, we ran \textsc{tphot} using FFT convolution and a cells-on-objects fitting configuration with the LU linear system solver. 

Since not all objects detected in the \textit{HST} map are also detected in the IRAC images, the exact fraction of IRAC flux recovered during the \textsc{tphot} fitting process is difficult to quantify. However, we recover $85-100\%$ of the original IRAC flux (measured using \textsc{Source Extractor}) for sources with clear singular IRAC counterparts (e.g. members 5 and 7). Additionally, visual inspection of the residual maps from our fitting procedure confirms no major systemic issues were generated during the convolution process (e.g. systemic offsets, shadows from inaccurate PSFs/kernels, black spots from spurious overestimated fluxes). Thus, we interpret our fits as successful and consider the resulting deblended fluxes as representative. 

We also explored whether astrometric offsets between the \textit{Spitzer} IRAC images and \textit{HST} images could affect our counterpart matching and deblending process. We searched for all \textit{HST} counterparts in a $1.6^{\prime\prime}$ radius (corresponding to the IRAC Channel 1 \textsc{fwhm}) from the IRAC sources and found an average offset of $\delta\textsc{RA} = 0.06\pm0.34^{\prime\prime}$ and $\delta\textsc{DEC} = 0.24\pm0.34^{\prime\prime}$ between matched counterparts, which is comparable to the \textit{HST} \textsc{fwhm} but significantly smaller than the IRAC \textsc{fwhm}. These offsets were not systematic in any direction. 

Since \textsc{tphot}-IRAC photometry is based on \textit{HST} coordinates, the IRAC fluxes are summed in a similar fashion: S/N\,$<2$ sources are used only as upper limits in the SED fitting process (see Section \ref{sec:SEDs}), and the remaining fluxes are summed to form a single measurement that's used in the SED modeling process, with uncertainties added in quadrature. 

\section{SED Modeling} \label{sec:SEDs}
We use the \textsc{cigale} \citep[Code Investigating GALaxy Emission,][]{cigale,cigale3,cigale2} SED modeling tool in \texttt{python} to generate SEDs for each of the 11 objects. \textsc{cigale} uses an energy balance principle based on conservation of energy between stellar emission, dust attenuation, and dust emission from UV to far-IR wavelengths, and estimates individual galaxy properties using a Bayesian approach (see \citet{cigale3} for full details). We select flux densities measured at signal-to-noise\,$\gtrsim2$ (listed in Table \ref{table:observations}). Detections with S/N\,$<2$ are used as upper limits in the SED fitting process, which are treated in the SED fitting process as described in detail by \citet{sawicki12} and \citet{cigale2}. We refer the reader to Section \ref{identifying} and \ref{sec:tphot} for details on multiwavelength counterpart selection. 

We use the following templates and modules to model each DRC member: a \citet{chabrier} IMF with a delayed star formation history; the \citet{bruz03} stellar population synthesis models; the \citet{calzetti00} starburst dust attenuation curve, and the \citet{draineli07, draine2014} two component dust emission models. We fit over a wide range of e-folding times ($5-200$\,Myr, given the age of the Universe at this redshift), metallicities ($0.0001-0.05$\,Z$_\odot$), and UV slopes ($\beta = -1.75-2$, \citet{caseyblue}). For the dust emission component, we allow the models to explore all PAH mass fractions available in the module, minimum diffuse dust radiation intensities, $U_\mathrm{min}$, of 0.1, 0.5, 1, 5, 10, or 50 (from adult stellar populations) combined with a fixed maximum radiation field intensity of 10$^7$ (from star-forming regions, \citet{draine2014}), a fixed mid-IR power-law slope of $2$ \citep{casey12}, and possible percentages of dust emission linked to star-forming regions (as opposed to ambient heating by adult stars) of 50, 75 and 100\%.

We present the best fit models in Figure \ref{fig:seds} and corresponding galaxy properties in Table \ref{table:properties} for all components. We do not include a SED model for DRC-4 as only upper limits are measured at $\lambda_\mathrm{obs} < 2$\,mm; still, we list SED-estimated properties for DRC-4 and caution against further interpretation of these values without further photometry to constrain them. For the remaining objects: all of the best-fit model SEDs have reduced $\chi^{2} < 2$, except for DRC-2 with $\chi^{2}_\mathrm{red} \approx 3$. For this object, the higher $\chi^{2}$ is likely due to the excess emission measured at observed-frame 3.6\,$\mu$m; DRC-2 is likely an ongoing major merger (see e.g. Figure \ref{fig:insetcontours}), and the excess 3.6\,$\mu$m emission may be driven by increased (and redshifted) H$\alpha$ flux from a recent extreme star-formation event, captured in the wide-banded IRAC Channel 1 \citep{smit16}. We also note that our resulting SFRs and infrared luminosities are similar to those found in Oteo et al., within 1$\sigma$ uncertainty.

While the reduced $\chi^{2}$ values are acceptable, we recognize that complex SED modeling techniques can be highly degenerate when there are more free parameters than data points to constrain them. This is particularly true for DSFGs as this population's stellar properties are not yet fully characterized. For example, while a stellar initial mass function (IMF) with more massive stars is favored for DSFGs \citep[e.g.][see also \citet{hayward2013b}]{baugh05, zhang18, cai19}, employing different IMFs, each weighted towards more massive stars, can result in a $2-3\times$ difference in stellar mass estimates \citep[e.g.][]{michalowski10,hainline,michalowski12}. Moreover, variations in star formation histories and stellar population synthesis models can further degenerate stellar mass estimates in DSFGs \citep[e.g.][]{hainline, michalowski14, wiklind14}. 

For this work, we can check the SED-derived stellar masses by comparing them to estimates based on rest-frame 1.6\,$\mu$m absolute magnitudes (observed-frame $\lambda = 8.0\,\mu$m), which is taken directly from respective best fit SEDs. This wavelength traces the stellar peak while also limiting the effects of dust extinction, as well as contributions from thermally pulsing asymptotic giant branch stars and/or AGN \citep{hainline09, chapman09, henriques11}. We derive an average $M_{H}$ of $-26.06\,\pm1.40$, which is in agreement (within uncertainty) of the SMGs studied in \citet{hainline09} and \citet{simpson14}. We apply the mass-to-light ratio $L_{H}/M_* = 7.9$, which was derived from a sample of SMGs in \citet{hainline09} and used for protocluster DSFGs in \citet{chapman09} and \citet{casey16}, deriving stellar masses that are within a factor of two of the SED-derived estimates. 

These similarities between stellar mass estimates could be driven by the unconstrained mid-IR portion of the SED that is red-ward of the observed frame $4.5\,\mu$m measurement. We do not have data to constrain the redder side of the rest-frame 1.6\,$\mu$m ($\lambda_{obs} = 8\,\mu$m) bump, and emphasize that further analysis and follow up observations are necessary to fully characterize these objects. Still, many other $z>2$ protoclusters in the literature that have optical/near-IR and far-IR measurements were analyzed using similar SED decomposition methods. Thus, we move forward using the SED-derived properties in this paper to put the DRC into context with outside literature. 

The possible presence of active galactic nuclei (AGN) embedded within a galaxy could also introduce additional uncertainties in the SED fitting process \citep[e.g.][]{edelson,murphy09,elbaz11,mullaney,ciesla}. AGN-warmed dust is shown to have the strongest contributions ($>30\%$) between rest-frame $\lambda = 1.0-30\,\mu$m \citep[e.g.][]{itme}, which could cause an overestimation of up to $60\%$ in stellar mass for an individual galaxy. However, this is less of an issue for SED-derived stellar masses within the DSFG population \citep{michalowski14}. If present, AGN contributions would likely have the most significant impact on DRC-6, the only galaxy for which \citet{oteo18} identifies radio emission in excess of the FIR-radio correlation and flat radio spectrum known for typical (i.e. non-active) DSFGs \citep{ibar10}. DRC-3 also exhibits an upturn in the mid-IR shown in the increasing $3.6\,\mu$m to $4.5\,\mu$m measured flux, which may also be indicative of a heated dust component from an obscured AGN. 

\begin{table}[t]
    \centering
    \caption{Galaxy properties derived from SED fitting.}
    \begin{tabular}{l @{\hspace{0.4\tabcolsep}} c @{\hspace{0.4\tabcolsep}} c @{\hspace{0.4\tabcolsep}} c @{\hspace{0.4\tabcolsep}} c @{\hspace{0.4\tabcolsep}} c c|c|c|c|c}
    \hline \hline \\
    ID & log(L$\mathrm{_{IR}}$) & SFR & log(M$_{*}$) & log(M$_\mathrm{H_{2}}$)\tablenotemark{a} & T$_\mathrm{d}$\tablenotemark{b}\\
     & [$\times 10^{12} \mathrm{L_\odot}$] & [M$_\odot$/yr] & [$\times 10^{10}$\,M$_\odot$] & [$\times 10^{11}$\,M$_\odot$] & [K]\\
    \\
    \hline \\
    DRC-1 & 19\,$\pm\,5$ & 1744\,$\pm\,1162$ & 16\,$\pm\,7$ & 8.62$\pm2.15$ & 35\\
    DRC-2 & 10\,$\pm\,8$ & 1132\,$\pm\,1013$ & 8\,$\pm\,6$ & 2.94$\pm0.73$ & 40\\
    DRC-3 & 18\,$\pm\,8$ & 1527\,$\pm\,1303$ & 17\,$\pm\,12$ & 2.68$\pm0.67$ & 42\\
    DRC-4\tablenotemark{c} & 4 & 200 & 16 & 1.41$\pm0.35$  & 28 \\
    DRC-5\tablenotemark{d} & 4\,$\pm\,11$ & 167\,$\pm\,375$ & 15\,$\pm\,8$ & 1.20$\pm0.30$ & 31\\
    DRC-6 & 2\,$\pm\,2$ & 190\,$\pm\,190$ & 3\,$\pm\,2$ & 1.14$\pm0.29$ & 21\\
    DRC-7 & 2\,$\pm\,6$ & 227\,$\pm\,303$ & 5\,$\pm\,5$ & 0.71$\pm0.18$ & 31\\
    DRC-8 & 5\,$\pm\,4$ & 394\,$\pm\,448$ & 6\,$\pm\,2$ & 0.22$\pm0.06$ & 56\\
    DRC-9 & 2\,$\pm\,2$ & 226\,$\pm\,281$ & 3\,$\pm\,2$ & 0.17$\pm0.04$ & 64\\
    DRC-10 & 1\,$\pm\,1$ & 60\,$\pm\,88$ & 2\,$\pm\,1$ & 0.16$\pm0.04$ & 40\\
    DRC-11 & 1\,$\pm\,1$ & 114\,$\pm\,142$ & 2\,$\pm\,1$ & 0.16$\pm0.04$ & 43\\
    Avg\tablenotemark{e} & 6\,$\pm\,6$ & 543\,$\pm\,586$ & 9$\pm\,6$ & 1.76$\pm2.36$ & 40$\pm\,12$\\
    \end{tabular}
    \tablenotetext{a}{Molecular gas masses are derived from converting 2\,mm flux densities to rest-frame 850\,$\mu$m luminosities. See Section \ref{sec:gas} for more details.}
    \tablenotetext{b}{Dust temperatures derived from SED fitting described in Section \ref{sec:SEDs}.}
    \tablenotetext{c}{All properties for DRC-4 are general estimates, based only on using upper limit near-IR photometry in SED fitting. See Section \ref{sec:SEDs} for more details.}
    \tablenotetext{d}{Properties derived assuming DRC-5 is at $z=4.002$ (not confirmed).} 
    \tablenotetext{e}{Averages do not include DRC-4.} 
    \label{table:properties}
\end{table}

\section{DRC Compared to Field Galaxies} \label{sec:indv}
When dissecting the individual properties of cluster versus field galaxies out to $z\sim2$, studies find weak evidence of distinguishable differences between the populations, often suggesting minor increases in the quiescent and/or quenched fraction and the massive galaxy population in overdense environments \citep[e.g.][]{koyama13,zavala19}. In the following sections, we discuss some differences we do (or don't) see in our protocluster core population of DSFGs when compared to $z\sim4$ field galaxies, and stress that more stringent conclusions could be drawn with additional optical/near-IR data. DRC-4 is not included in this analysis, making our focus on only ten of the DRC components (one of which, DRC-5, is not yet spectroscopically confirmed at $z=4.002$).  

\vspace{-7pt}

\begin{figure*}[ht]
\begin{center}
\includegraphics[trim=0cm 0cm 0cm 0cm, scale=0.5]{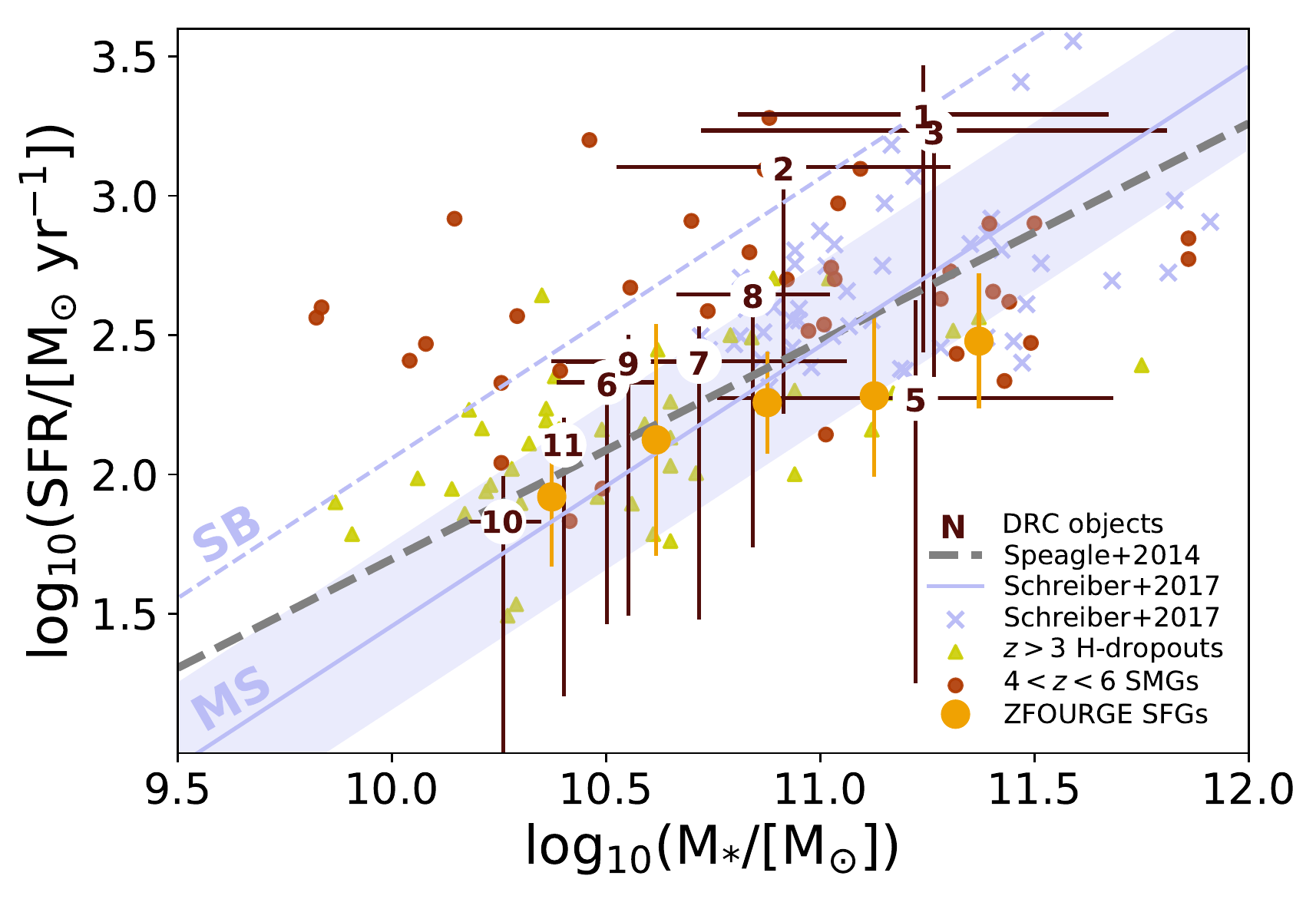}
\includegraphics[trim=0cm 0cm 0cm 0cm, scale=0.5]{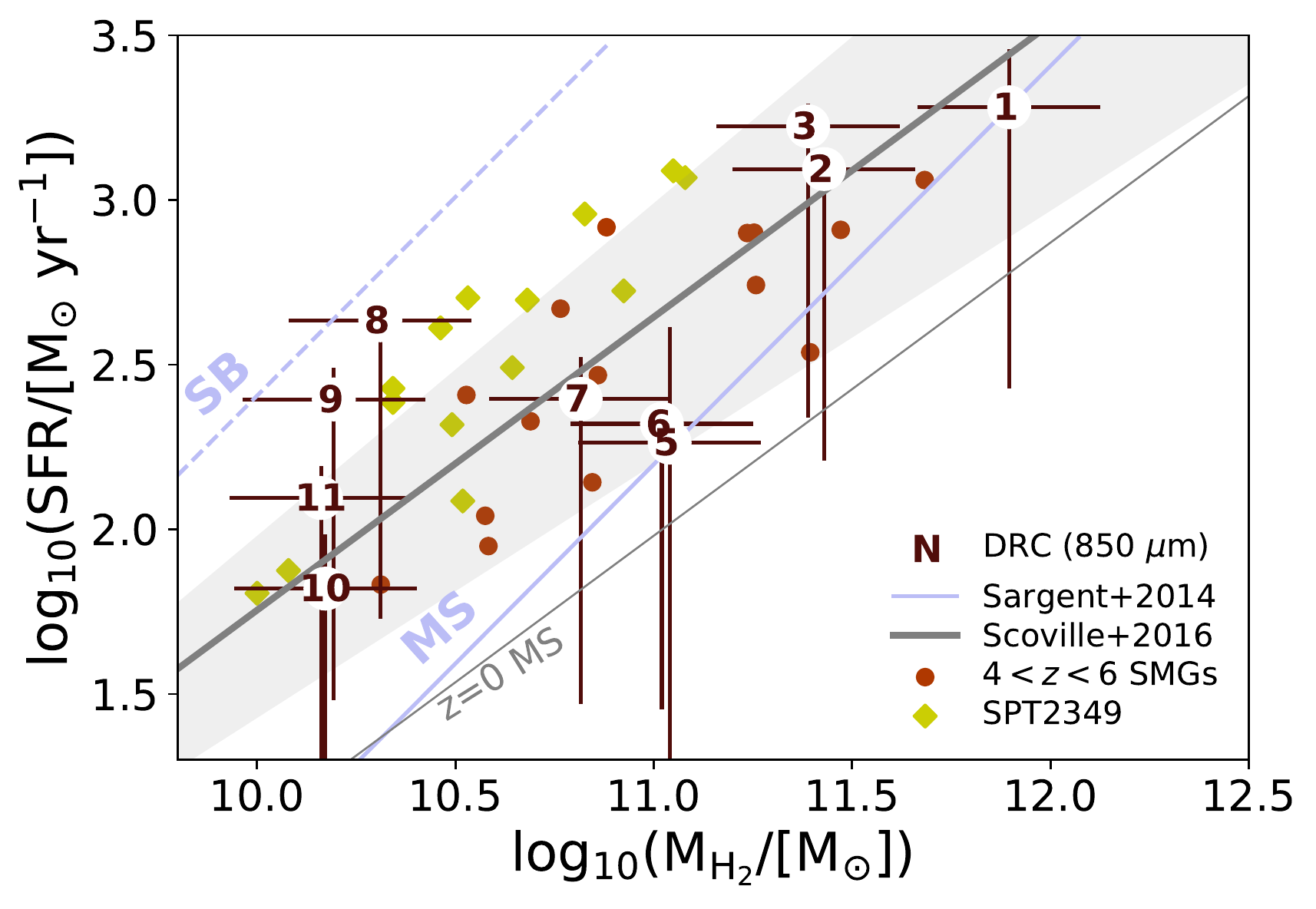}
\end{center}
\caption{
\footnotesize \textit{Left}: The SFR--M$_*$ relationship at $z\sim4$. All DRC components (except DRC-4, see Section \ref{sec:SEDs}) are represented by their IDs. The orange points represent the average $3<z<4$ main-sequence found in \citet{zfourge}. Red points are $z\gtrsim4$ SMGs taken from \citet{fujimoto17} and \citet{scoville}. Green triangles represent the massive star-forming $z>3$ $H$-band dropout population from \citet{wang19}. Blue x's represent the ALMA Redshift 4 Survey of massive (M$_* > 5\times10^10$\,M$_\odot$) star-forming galaxies \citep{sch17}. The blue and grey lines represent the $z=4$ main-sequence derived in \citet{sch17} and \citet{speagle}, respectively, with the blue shaded region corresponding to the 1$\sigma$ uncertainty from the Schrieber et al. sample, and the dashed line corresponding to the starburst region $4\times$ above the main-sequence. \textit{Right}: Molecular gas mass versus star formation rate. Green triangles are from the \citet{miller18} $z=4.3$ sub-mm protocluster, red points are from the \citet{scoville} $<z>\,=4.4$ ALMA continuum sample, and the gray line marks the star formation law at $z=4$ generated by Equation (2) in \citet{scoville}. Redshift independent main sequence (solid) and starburst (dashed) relationships from \citet{sargent14} are in blue.
}
\label{fig:mainseq}
\end{figure*}

\subsection{Main Sequence Evolution} \label{sec:msevo}
In Figure \ref{fig:mainseq} left, we compare the SFRs and stellar masses for each DRC member to other known $z\sim4$ populations. We include a sample of mass complete (M$_{*} \gtrsim 1.6 \times 10^{10}$\,M$_\odot$) $3<z<4$ main-sequence star-forming galaxies from the ZFOURGE survey \citep{zfourge}, massive dusty $3<z<6$ star-forming galaxies from \citet{wang19}, a SFR-limited ($\ge 100$\,M$_{\odot}$\,yr$^{-1}$) sample of $4<z<6$ ALMA observed SMGs with \textit{HST} counterparts \citep{fujimoto17}, and a population of magnification-corrected gravitationally lensed SMGs at similar redshifts from \citet{scoville}. 
    
In general, all cluster core members are massive, averaging at $9\pm6\times10^{10}$\,M$_\odot$, and reside within 1$\sigma$ of the $z\sim4$ \citet{sch17} and \citet{zfourge} star-forming main sequences. Nine out of ten of the protocluster members in discussion are likely larger than $\sim3\times10^{10}$ solar masses, while only $64/654$ (or $\sim10\%$) of $z>3$ ZFOURGE galaxies achieve such high stellar mass \citep[][and the true $z=4$ percentage is likely even less]{zfourge}. In the higher-redshift SMG samples, we find a much higher fraction of massive galaxies \citep[$\sim70\%$, ][]{scoville, fujimoto17}, which may be more representative of the DRC SMGs. While some $z=2$ studies report high fractions of massive galaxies in overdense environments \citep[e.g.][]{koyama13}, such a high fraction in the DRC may be driven by selection bias towards massive and bursty systems; further followup observations searching for nearby normal star-forming galaxies are required to substantiate this claim. 
    
Stellar mass functions of far-IR bright star-forming galaxies at $z\sim4$ estimate a number density of $n\sim 10^{-3.6}$\,Mpc$^3$ for objects $>3\times10^{10}$\,M$_\odot$ \citep{sch15}; in a $260\,\mathrm{kpc}\times310\,\mathrm{kpc}\times87\,\mathrm{Mpc} \approx 7$\,Mpc$^3$ comoving volume like this protocluster, we expect to see 0.00049 galaxies as massive as each of the DRC members. This corresponds to a galaxy overdensity of $\delta_\mathrm{gal} = (8-0.00049)/0.00049 > 10,000\times$ the field density. While this value may decrease once more protocluster members are confirmed in a wider volume, it underpins the evolutionary concept outlined in \citet{casey16} where overdensities of rare and massive DSFGs are likely correlated, not serendipitous, with massive protocluster evolution.
    
Protocluster members closest to the starburst\footnote{In this work, starbursts are defined as having SFRs that are $4\times$ greater than the main sequence at a given stellar mass \citep{rodighiero}.} region, components 1-3, morphologically exhibit possible interactions or ongoing mergers in the rest-frame UV, which some studies argue is a primary driver of a galaxy's presumed short-lived starburst phase \citep[e.g.][]{sanders88}. We also note that a part of the \citet{sch17} sample occupies a similar high-mass near-bursty region of the SFR-M$_*$ plane. This sample is a closer evolutionary proxy to the DRC, focusing strictly on massive ($>5\times10^{11}$\,M$_\odot$) \textit{HST} $H$-band selected and ALMA observed star-forming galaxies at $3.5<z<4.7$. However, only two of these sources were identified as undergoing close ($<1^{\prime\prime}$) interactions, with additional environmental information currently unavailable. Moreover, there exist several DRC components that are on or below main-sequence with disturbed or merging rest-frame UV morphologies. Thus, on the SFR-$\mathrm{M_*}$ plane, it is unclear whether merging activity in overdense regions creates starburst galaxies. 
    
Unraveling any inherent differences in this protocluster core versus field populations, such as an increased fraction of massive galaxies and/or starburst activity, requires further observations. Additional observations in the rest-frame UV/optical could establish the presence of normal star-forming galaxies, galaxies with post-starburst signatures, as well as quiescent early-type galaxies (which we are now seeing out to $z\sim3.5$ in the field, e.g. \citet{forrest19}) -- all of which would pose significant implications on the galaxy growth and evolution in overdense environments. Precise spectroscopic redshift information on these additional populations would also constrain the impact of filamentary dynamics on galaxies in early protoclusters. 

\subsection{Gas Properties} \label{sec:gas}
\begin{figure}
\begin{center}
\includegraphics[trim=0cm 0cm 0cm 0cm, scale=0.33, width=.45\textwidth]{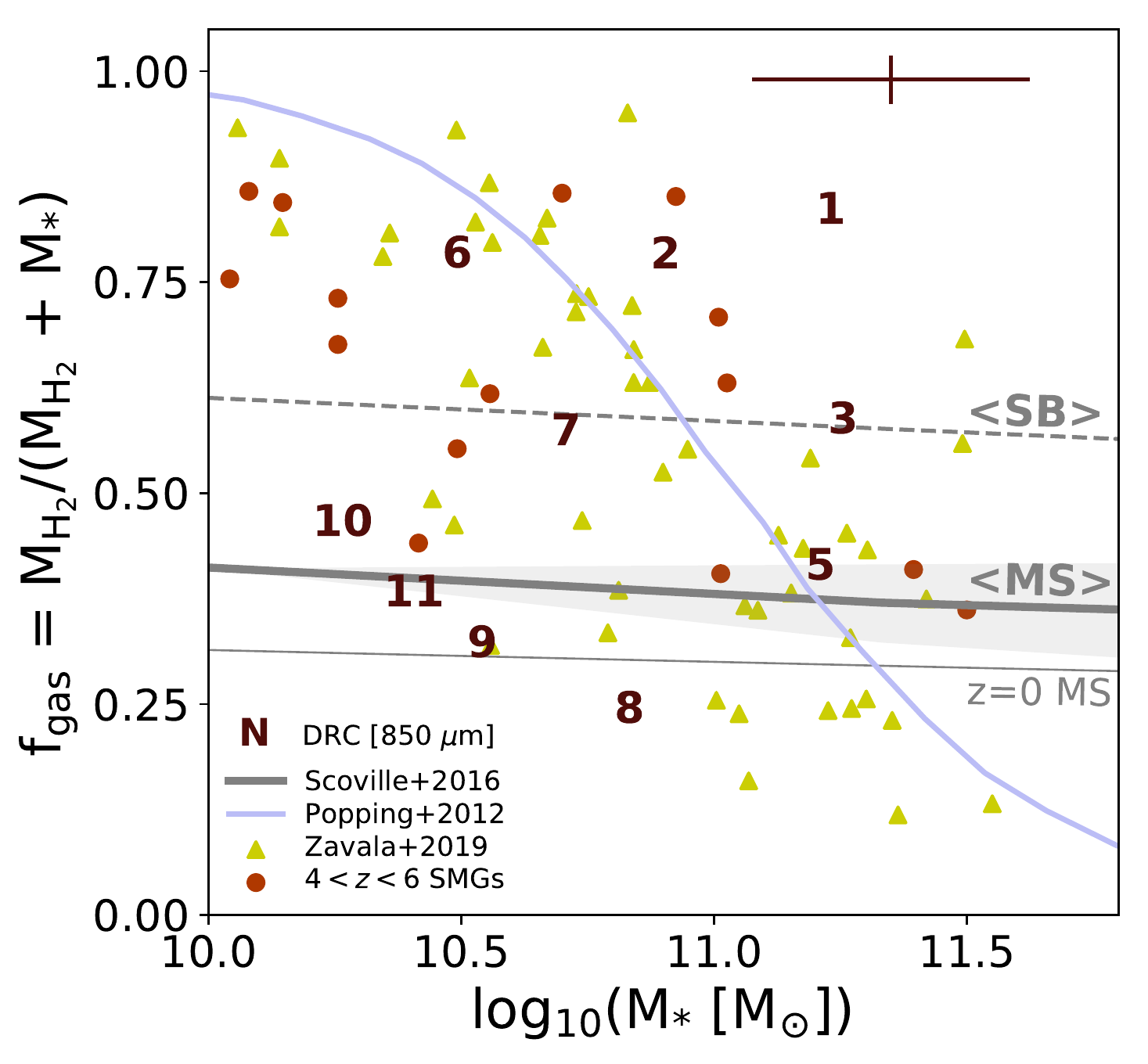}
\end{center}
\caption{
\footnotesize Molecular gas mass fraction as a function of stellar mass. Numbers mark the individual DRC components. The average 1$\sigma$ uncertainty for DRC components is provided in the top right corner. The blue curve represents the $\langle z \rangle=2$ relationship derived in \citet{popping12}, where high mass galaxies are expected to have lower gas mass fractions (due to both gas depletion and halo shock heating). Red points mark $z\sim4$ field SMGs \citep{scoville}, and green triangles mark $z=2-3$ ALMA-detected protocluster galaxies \citep{lee17,gg19,tadaki19,zavala19}. Solid and dashed grey lines represent the average gas mass fraction from \citet{scoville} for main-sequence and starburst galaxies at $z\sim4$; we also show the average for main-sequence galaxies at $z=0$ evolved from the \cite{scoville} relationship.}
\label{fig:gas}
\end{figure}

Cluster environments as global systems are expected to have massive intracluster reservoirs of gas. Yet, at the individual galaxy level, some studies show that there is little to no change in gas mass fractions when considering galaxy environments out to $z\sim2.5$ (e.g. \citet{darvish18} and \citet{zavala19}, see also \citet{tadaki19} and \citet{noble17}). In Figures \ref{fig:mainseq} (right) and \ref{fig:gas}, we explore whether this holds for DRC galaxies.

We derive molecular gas masses using the method outlined in \citet{scoville}. This method is built on the observed and theoretical link between the Rayleigh-Jeans tail that traces dust emission and the molecular gas within the ISM of SMGs; and, it is calibrated using the ratio between rest-frame 850\,$\mu$m luminosity (L$_{850\,\mu m}$) and molecular gas mass (M$_\mathrm{H_2}$). This ratio, aka $\alpha_{\mathrm{850 \mu m}}$, absorbs inherent variations in dust temperature, opacities, and abundances, and was further calibrated using CO\,(1-0) measurements in DSFGs. We use the value given in Scoville et al. where $\alpha_{\mathrm{850 \mu m}} = 6.7\pm1.7 \times 10^{19}\,$erg\,s$^{-1}$\,Hz$^{-1}$\,M$_\odot^{-1}$. Considering possible variations in true galaxy dust temperature, gas mass uncertainties using this method are estimated at $\sim25\%$. We refer the reader to Appendix A of \citet{scoville} for further details on derivation and resulting uncertainties.

For each individual galaxy, we convert observed-frame 2\,mm flux densities ($\lambda_\mathrm{rest}=400\,\mu$m) to molecular gas masses using the following equation:

\begin{align*}
    \mathrm{M_{H_2}} = 1.78\,\mathrm{S_{\nu_{obs}} [mJy]}\,(1+z)^{-4.8} \\
    \times \Big\{\frac{\nu_\mathrm{850 \mu m}}{\nu_{obs}}\Big\}^{3.8} (D_L\,[\mathrm{Gpc}])^2 \\ \times \Big\{\frac{6.9\times10^{19}}{\alpha_\mathrm{850 \mu m}}\Big\}\frac{\Gamma_\mathrm{RJ}}{\Gamma_{\mathrm{RJ},\,z=0}}\,10^{10}\,\mathrm{M_\odot} \stepcounter{equation}\tag{\theequation}\label{eqn:mgas}
\end{align*}

\noindent where $\mathrm{S_{\nu_{obs}}}$ is the observed flux density at $\lambda_\mathrm{rest}>250\,\mu$m (where the dust is considered optically thin), $\nu_{obs}$ is the frequency of the observed flux density ($=150$\,GHz), and $D_L$ is the luminosity distance at $z=4.002$. $\Gamma_\mathrm{RJ}$ is the Rayleigh-Jeans (RJ) correction factor for deviation from the rest-frame Planck function (i.e. $B_{\nu_\mathrm{rest}}/\mathrm{RJ}_{\nu_\mathrm{rest}}$), developed in \citet{scoville} and given by

\begin{align*}
    \Gamma_\mathrm{RJ}(\nu,T_d,z) = \frac{h\nu_{obs}(1+z)/kT_d}{e^{h\nu_{obs}(1+z)/kT_d}-1}
    \stepcounter{equation}\tag{\theequation}\label{eqn:gamma}
\end{align*}

\noindent where \textit{h} is the Planck constant, \textit{k} is the Boltzmann constant, and $T_d$ is the galaxy's mass-weighted dust temperature (assumed to be 25\,K to be consistent with other work). Using this approach, we estimate molecular gas masses at $0.16-8.6\times10^{11}$\,M$_\odot$, with an average M$_\mathrm{H_2} = 1.76\pm2.36\times10^{11}$\,M$_\odot$ (see Table \ref{table:properties}). While a cooler dust temperature is possible for DSFGs \citep[e.g. 15\,K][]{hwang10}, it is unlikely the case for the DRC since the temperature of the CMB at this redshift is $\sim13.5$\,K. Using a hotter dust temperature, such as the individual temperatures determined in the SED fitting process (e.g. 40\,K, see Table \ref{table:properties}), results in only a marginal decrease in molecular gas mass estimates (by $\sim10-20\%$, or $\sim0.1-0.4$\,dex). 

For objects 1-4, the values derived using both $T_d = 25\,$ and the SED-derived dust temperatures are within $\sim1$\,dex of the gas masses derived using [C\,\textsc{i}](1-0) line emission in \citet{oteo18}. Due to possible degeneracies in SED-derived dust temperatures and to remain consistent with outside literature, we move forward in this analysis using the gas masses estimated with a mass-weighted dust temperature of 25\,K.

Objects 1-6 also have 3\,mm observations, which is closer to the rest-frame 850\,$\mu$m ($\lambda_\mathrm{obs} = 4.25\,\mu$m) emission used to derive the Scoville et al. relationship. Under the same assumptions listed above for the 2\,mm data, we determine 3\,mm gas estimates for DRC objects 3, 5, and 6 that are slightly larger (by 0.06, 0.14, and 0.31 dex, respectively); for objects 1, 2, and 4 the 3\,mm gas estimates are smaller (by 0.17, 0.12, and 0.05 dex, respectively). With the exception of DRC-6, the differences between the 2 and 3\,mm estimates are generally within the included 1$\sigma$ uncertainties on the 2\,mm estimates. Since 2\,mm data is available for all sources, and the differences between the two mass estimates are marginal, we choose to use the 2\,mm-derived gas masses (over the 3\,mm) throughout this work. 

With the above method, we avoid the major uncertainties that come with assumed gas mass estimates from CO SLED analysis at high-\textit{J} transitions. High-\textit{J} transitions, like the  $J=6\rightarrow5$ transition line detected in DRC objects, trace denser regions of gas than the lower \textit{J} transitions, which trace cooler, diffuse gas reservoirs throughout the galaxy. Still, as a comparison to our luminosity-derived gas masses, we derive line-driven gas masses using the $^{12}$CO(6-5) luminosities provided in Oteo et al.

Assuming that DRC objects have similar spectral line energy distributions (SLEDs) as other high-z SMGs, we can use the $^{12}$CO(6-5) line luminosity to convert to the ground-state $^{12}$CO(1-0) luminosity, as tabulated in \citet{bothwell2013}. We assume a CO-to-H$_2$ conversion factor of $\alpha_\mathrm{CO} = 1.0$\,M$_\odot$\,(K\,km\,s$^{-1}$\,pc$^2$)$^{-1}$, as used in Bothwell et al. and others for SMGs \citep[e.g.][]{tacconi08}, and determine gas masses of M$_\mathrm{H_2} = 0.1-6\times10^{10}$\,M$_\odot$, about an order of magnitude smaller than the masses derived using the dust continuum tracer. 

In Figure \ref{fig:mainseq}, we show DRC members on the SFR-M$_\mathrm{H_2}$ plane using the molecular gas masses derived with the 2\,mm flux densities. Eight out of ten members lie within the main-sequence regime with total gas masses $>10^{10}\,$M$_\odot$ -- estimates that are similar in mass and spread to the \cite{scoville} DSFG sample and the similarly compact and star-forming $z=4.3$ SPT-2349 protocluster of DSFGs \citep{miller18, hill2020}. About 50\% of our sample have relatively large gas masses at $>10^{11}\,$M$_\odot$, while the same is only true for none of the SPT protocluster (core) members and 6/15 of the $z\ge4$ SMG sample. DRC-8 and 9 have elevated SFRs near the starburst regime (above the expected $z\sim4$ main-sequence star-formation law at a given molecular gas mass \citep{scoville}). These objects also lie within the M$_*$-SFR main-sequence which may suggest that their high star-forming efficiencies (=SFR/M$_\mathrm{H_2}$) are driven by relatively small gas reservoirs rather than extreme rates of star-formation.

Assuming a closed box scenario with a constant star-formation rate and no major influx of cold gas, we can estimate individual gas depletion timescales, $\tau_\mathrm{depl} = \mathrm{M_{H_2}}/\mathrm{SFR}$. Of course, in overdense regions like these, mergers and fresh gas inflows are expected, but we can still use the instantaneously measured gas depletion times to understand the efficiency at which these objects are turning gas into stars at this given moment (while also neglecting any impacts from feedback).

Despite their large gas reservoirs, DRC objects will deplete their gas in an average of $\sim260\pm180$\,Myr, which is similar to the mean $\tau_\mathrm{depl}$ for the SPT $z=4.3$ protocluster \citep{miller18} and the $z>4$ field SMG sample \citep{scoville}, at 122$\pm53$ and 300$\pm160$\,Myr, respectively. DRC gas depletion timescales are more consistent with general field SMG gas surveys at high-z \citep[$\sim100$\,Myr, e.g.][]{tacconi08,aravena16,yang17} than those of local interacting infrared luminous galaxies \citep[e.g.][]{sanders86}. If we assume that no major gas is flowing in to support these SFRs, these timescales may indicate that these objects will deplete their gas reservoirs by $z\sim3$. 

Dividing stellar mass by the star-formation rate, we can estimate the stellar-mass build up timescale (assuming that these SFRs have been sustained in the past); we derive build up timescales ranging from 70-300\,Myr with a median of $\sim$160\,Myr -- which is within the expected lifespan of the starbursting phase for submillimeter galaxies \citep[e.g.][]{narayanan10}. Still, in deep potential wells like that of this overdensity, gas is expected to flow in at increased rates, which may actually sustain these extreme bouts of star-formation for longer periods of time. Further deep observations for extended, cold gas surrounding the protocluster would be necessary to confirm this latter scenario.

An additional metric we can inspect is the gas mass fraction, f$_\mathrm{gas}$ = M$_\mathrm{H_{2}}$/(M$_\mathrm{H_{2}}$ + M$_{*}$), which is expected to decrease with increasing stellar mass \cite[e.g.][]{popping12, genzel15}. In Figure \ref{fig:gas}, we see that this appears to be the case for the \citet{scoville} $z\sim4$ field SMGs and other $z=2-3$ protoclusters \citep{zavala19,tadaki19,gg19} -- as well as for DRC members. DRC members span a wide range of gas fractions, from 25-80\%, with an average f$_\mathrm{gas}=52\pm20\%$, across all galaxy stellar masses. Objects 8 and 9 have some of the smallest gas reservoirs ($<3\times10^{10}$\,M$_\odot$), are within the  SFR-M$_*$ main-sequence law, and also lie at the bottom edge of the f$_\mathrm{gas}$-M$_{*}$ expected relationship. With gas depletion timescales of $\sim50$\,Myr, it is possible that these objects are much closer to depleting their gas supplies than the other core members, and on their way to becoming some of massive quiescent galaxies that dominate in cluster cores by $z\sim3.5$ \citep[e.g.][]{del07,ito19}.

Understanding the growth and quenching of gas reservoirs in overdense environments at $z>3$ requires additional follow-up observations in the rest-frame far-IR/sub-mm. While we show that the population of SMGs in a protocluster spans a wide range of gas-richness, we cannot draw further conclusions on whether specific quenching (or enhancement) activity is driven by environment until there are additional observations of other protocluster members, both within the DRC and other $z>3$ overdensities. With follow up dust continuum surveys of these overdensities, we can further constrain overarching questions in early cluster evolution, such as: How early does extreme stellar mass build up cease for brightest cluster galaxy progenitors? Do the majority of galaxies in overdense environments go through a starburst phase that's sustained with cold gas flows and, if so, at what point would virial shock heating disrupt these flows? 

\begin{figure}
\begin{center}
\includegraphics[trim=0cm 0cm 0cm 0cm, width=0.47\textwidth]{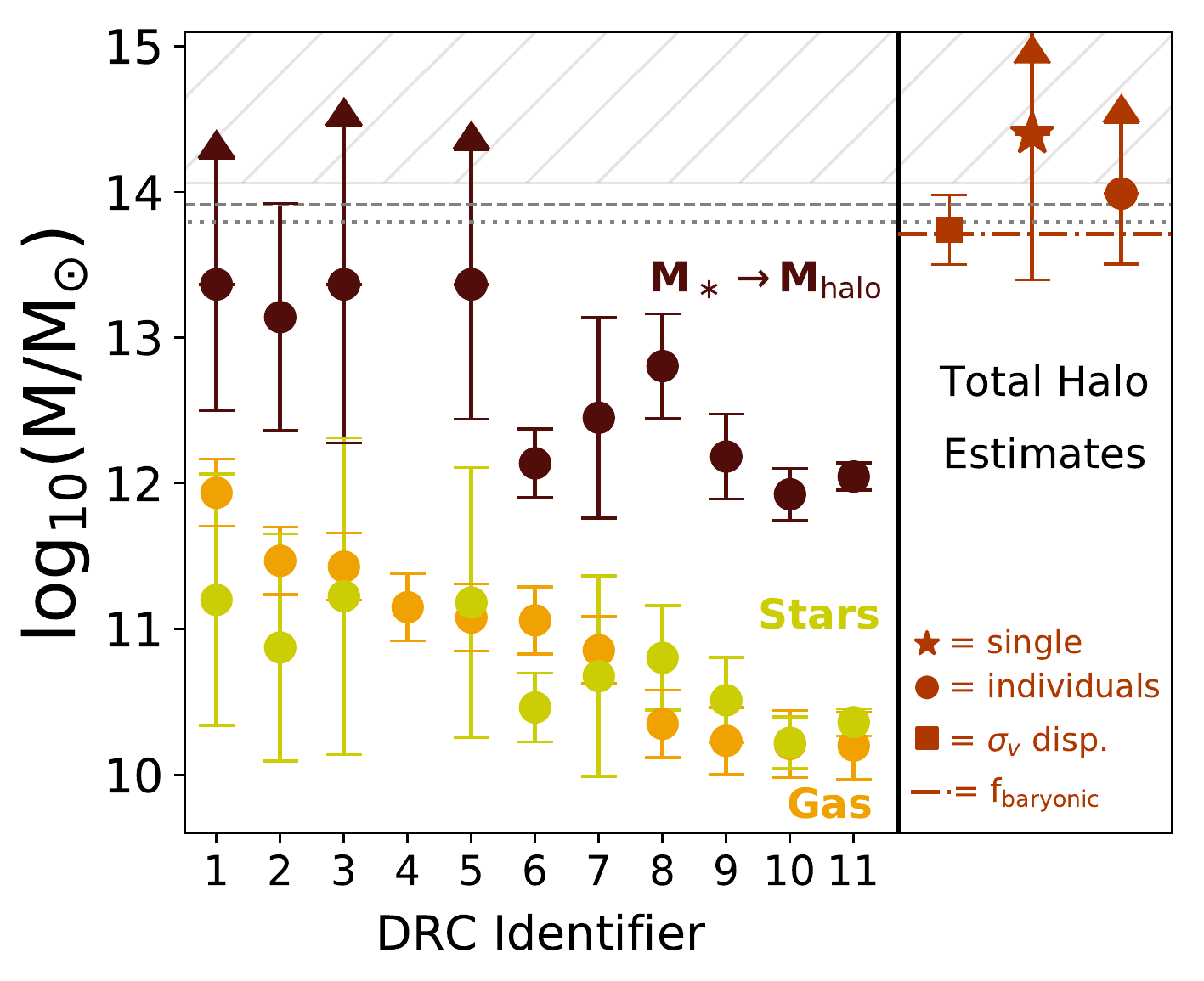}
\end{center}
\caption{
\footnotesize Halo mass estimates for the DRC. On the left, we show mass estimates for stars, gas, and total halos for each individual DRC component (gas only for DRC-4, see Section \ref{sec:SEDs}). The orange corresponds to the gas mass estimates using the rest-frame 850\,$\mu$m technique as outlined in \citet{scoville} and Section \ref{sec:gas}. The green corresponds to the stellar masses estimated using SED fitting, as outlined in Section \ref{sec:SEDs}. The dark red represents the individual total halo masses derived using stellar abundance matching at $z=4$ from \citet{behroozi}. The dotted, dashed, and hatched regions represent the 1, 2, and 3$\sigma$ exclusion curves for the most massive halos expected to be observable at $z=4$ in the $\sim600$\,deg$^2$ \textit{H}-ATLAS survey \citep{hh13}. On the right, we show total halo mass for the entire DRC, estimated using various methods. The square and corresponding error bars represents the range of halo masses found in \citet{oteo18} using velocity dispersion methods. The circle represents the sum of the halos from each individual DRC component as seen in dark red on the left. For the star: all DRC objects are treated as subhalos within one single overarching halo; their individual stellar masses are summed into a singular massive stellar component that's then used to reverse engineer halo mass estimates using stellar-to-halo mass ratios from \citet{behroozi}. Finally, the dark red dash-dotted line represents the halo mass assuming a baryonic-to-halo mass fraction of 5\%. 
}
\label{fig:halos}
\end{figure} 

\section{Cluster Halo Mass at $\lowercase{z}=4$} \label{sec:dmh}

\begin{figure}
\begin{center}
\includegraphics[trim=0cm 0cm 0cm 0cm, width=0.47\textwidth]{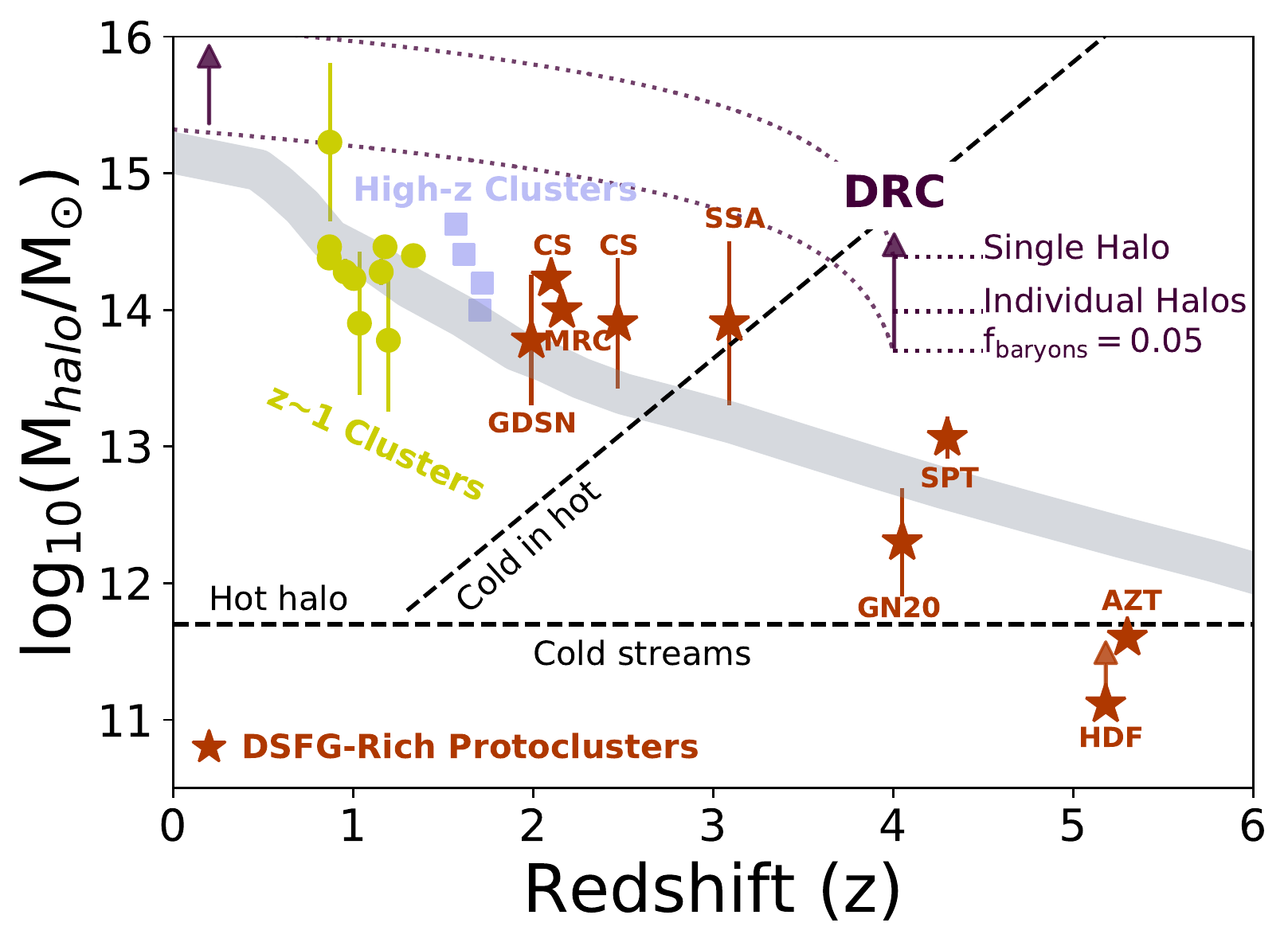}
\end{center}
\caption{
\footnotesize The halo mass evolution of protoclusters. For the DRC, we show the halo mass estimates from two methods in purple (presented in Section \ref{sec:dmh}): (i) generating halo masses using stellar-to-halo abundance matching \citep{behroozi} at the individual galaxy level (second highest estimate), (ii) the total halo mass if we combine all DRC stellar mass into a single component and use a range of stellar-to-halo mass ratios (most massive estimate), and (iii) the halo mass estimate found using the baryonic-to-dark matter ratio of 5\% (least massive estimate). The purple curves from $z=4$ to $z=0$ show the predicted evolution of halo growth for a \textit{single} halo in the Millenium and Millenium-II simulations \citep{mcbride09,fakma10}; we show the evolution for each of the DRC estimates. We consider these estimates as lower limits in this analysis as there are likely additional protocluster members not captured in the ALMA data. Green dots represent $z\sim1$ galaxy clusters from the GCLASS survey \citep{vander14}; blue squares represent $z=1-2$ virialized clusters \citep{newman,stanford,zeimann,mantz}; red stars represent SMG-rich $\gtrsim2$ protoclusters: the GOODS-N $z=1.99$ protocluster rich with AGN and SMGs, the COSMOS $z=2.10$ and $2.47$ protoclusters, the MRC1138256 aka `Spiderweb' protocluster at $z=2.16$, the SSA22 $z=3$ AGN and DSFG rich protocluster, the $z>4$ GN20, AzTEC-3, and HDF\,850.1 overdensities (with 1-3 DSFGs each), as well as SPT2349-56, the $z=4.3$ protocluster of 14 SMGs \citep{steidel98,kurk00,blain04,chapman09,daddi09,tamura09,lehmer09,capak11,kuiper11,walter12,hodge13,yuan14,dann14,casey15,diener15,chiang15,um15, casey16,miller18, hill2020}. The grey line shows the expected halo mass evolution of a Coma-like cluster \citep{chiang13}. The black dashed lines mark the different regions of gas inflow and cooling mechanisms on massive halos; gas inflows onto halos above the critical shock heating halo mass at $\sim10^{12}$\,M$_\odot$ are shock heated and thus the galaxy within is strangulated \citep{dekel06}. Galaxies in the ``cold in hot" regime may have penetrating cold gas flows that help sustain growth.
}
\label{fig:dmh}
\end{figure}

Weighing a high-$z$ protocluster requires a variety of assumptions. Typical methods used to derive galaxy cluster masses (e.g. measuring X-ray emission from the super-heated intracluster medium (ICM), or tracing Sunyaev-Zeldovich distortions on the CMB) are unavailable for objects like the DRC as most of these methods are fine-tuned for nearly or fully virialized $z\lesssim1$ clusters with an ICM. \citet{oteo18} attempt to overcome this by combining the velocity dispersion method \citep{evrard08} with ALMA $^{12}$CO($6-5$) line velocities to estimate a total DRC halo mass of $3-9 \times 10^{13}$\,M$_{\odot}$. This method requires an assumption that the DRC is already virialized. However, $z>3$ protoclusters exhibit generally aspherical mass distributions with large effective radii that vary based on the chosen line of sight \citep[e.g.][]{lovell, chiang17}; this is because eventual $z\sim0$ cluster members are tens to hundreds of Mpc apart at $z>3$. In the following, we weigh the DRC using three different methods, each of which comes with it's own assumptions and uncertainties. We present these estimates in Figure \ref{fig:halos} and \ref{fig:dmh}. As in the previous sections, we do not include DRC-4 in any calculations as we do not have reliable stellar mass estimates. 

First, we derive a modest estimate of the total cluster halo mass by summing the halo masses of each individual galaxy. This estimate requires the assumption that individual galaxy halos are closer to virialization than the protocluster itself, and that each galaxy formed it's own halo prior to coalescing in this overdense region. We use the stellar-to-halo abundance matching relationship in \citet{behroozi}, which is developed assuming that the bulk of baryonic mass in dark matter halos is tied up in adult stars, and that massive galaxies trace massive halos. We note that the \citet{behroozi} $z=4$ relationship does not extend into the stellar mass range we observe for the DRC, and thus those objects with stellar masses greater than $>10^{11}$\,M$_{\odot}$ are placed at the fixed maximum value of M$_\mathrm{halo} = 2\times10^{13}$\,M$_{\odot}$. This is applied to DRC objects 1, 3, and 5 - all three of which have stellar masses within uncertainty of the Behroozi et al. most massive halo bin. 

We determine individual halo masses of M$_\mathrm{DM} = 1-12\times10^{12}$\,M$_{\odot}$, with an average halo mass of $8\pm4\times10^{12}$\,M$_{\odot}$ (Figure \ref{fig:halos}, left), similar to other average DSFG halo masses seen overdense environments \citep[e.g.][]{hall18}. Summing up the individual components translates to a total cluster halo mass of $5\pm2\times10^{13}$\,M$_{\odot}$ (Figure \ref{fig:halos}, right). Errors are determined from uncertainties in stellar mass. This estimate agrees with the those derived in \citet{oteo18}. We note that if we do not include DRC-5 in this estimate (as it is not a spectroscopically confirmed member), the total halo mass drops to $\sim4\times10^{13}$\,M$_{\odot}$ -- roughly $<20\%$ less massive.

If we instead assume that these galaxies are (and maybe always have been) sitting and growing in the same halo, then the previous method would likely be an overestimate that ``double counts" dark matter mass in overlapping halos. Assuming each of these galaxies is close enough to be occupying one single massive halo (which, according to velocity space, may be true for 8/10 objects), one might sum all stellar masses into a single total stellar mass for the halo and then interpolate that value over the \citet{behroozi} relationship. Unfortunately, this total stellar mass goes well beyond the established $z=4$ Behroozi et al. relationship. Thus, to conceptualize this estimate, we instead use the stellar to halo mass ratios in Behroozi et al. (Fig. 7) for the largest $z=4$ halo mass value ($\sim10^{13}$\,M$_{\odot}$), which is set to M$_*$/M$_\mathrm{halo} = 0.003$. Combining this value with the total stellar mass for the DRC, we reverse estimate the cluster halo mass at $>3\times10^{14}$\,M$_{\odot}$. This is $\sim0.5$\,dex larger than the previous estimates, and the most massive estimate in this study. The lower limit of this method assumes a smaller halo and a more efficient stellar to halo mass ratio of M$_*$/M$_\mathrm{halo} = 0.02$ - which results in a cluster halo mass of $4\times10^{13}$\,M$_{\odot}$, a value similar to that of the most massive \textit{individual} galaxy halos. If DRC-5 is not included in either estimate, the total halo mass drops by about 30$\%$.

Finally, if we instead assumed a generous fixed baryonic-to-dark matter fraction of 5\% \citep[e.g.][]{behroozi18}, summing all stellar and gaseous components, we estimate a halo mass of $5\times10^{13}$\,M$_{\odot}$ -- similar to the individual halo mass estimate determined above, as well as the calculation from Oteo et al. 

In Figure \ref{fig:halos}, we compare these estimates to the 1, 2, and 3$\sigma$ exclusion curves for how likely a massive halo is to exist at $z=4$ in $\Lambda$CDM cosmology, as derived in \citet{hh13}\footnote{We determine these curves / statistics using the publicly available code from \citet{hh13} at: https://bitbucket.org/itrharrison/hh13-cluster-rareness/src/master/.} --  i.e. these exclusion curves mark the most massive clusters possible at 68, 95, and 99.7\% likelihood within the \textit{H}-ATLAS survey region of $\sim600$\,deg$^2$, with 1, 2, and 3$\sigma$ corresponding to upper mass limits of 6, 8, and 12$\times10^{13}$\,M$_\odot$, respectively.

Within uncertainty, each of the total halo mass estimates for the DRC do not necessarily break the 3$\sigma$\,(99.7\%) exclusion curve -- i.e. our current data portrays a massive structure that is rare in the Universe, but not improbable. However, we argue for a variety of reasons that our understanding of the DRC's true weight is incomplete (and therefore likely underestimated). Firstly, at large scales, these methods do not account for other additional cluster members that have yet to be detected, such as normal star-forming galaxies, post-starburst, and/or quiescent galaxies. We emphasize the impact of this point: the majority of galaxies in DSFG-rich $z<3$ protoclusters are normal star-forming systems \citep[$\sim85\%$, Table 1 in][]{casey19}; normal star-forming galaxies contain the majority of the cosmic stellar mass budget \citep{baugh05, rodighiero, sargent12}, and are found in large numbers in protoclusters out to $z\sim6$ \citep[e.g.][]{harikane19}. Thus, the presence of up to 10$\times$ as many normal SFGs as DSFGs in this protocluster core would have significant impact on the mass estimates, and therefore rarity, of the DRC. 

At smaller scales, the abundance matching method is developed on the basis that the most massive component in all halos is the stars; while the DRC presents gas-poor galaxies, it is still possible for gas-rich members to exist throughout the structure but outside of the ALMA field-of-view of this cluster core. Moreover, at high-$z$, it is possible for large gas reservoirs to become significant (or even the \textit{main} baryonic) contributors to the overall mass budget \citep[e.g.][]{casey19}. 

In Figure \ref{fig:dmh}, we compare DRC total halo mass estimates to that of other known protoclusters over a wide range of redshifts. At it's lowest estimate, it is already equally as massive as $z=2-3$ protoclusters \citep[most of which have halo mass estimates using the same stellar-to-halo matching technique at the individual galaxy level, ][]{casey16}, and at its largest the DRC is nearly as massive as $z=1$ virialized clusters \citep{stanford,zeimann,newman,mantz}. Again, we emphasize that additional, non-negligible mass is likely missing from the DRC in the form of less star-forming galaxies or other far-IR bright sources not within the original ALMA field-of-view - as found to be true for the SMG-rich SPT2349-56 protocluster at $z=4.3$ \citep{miller18}. Follow up ALMA spectroscopic scans on far-IR bright regions surrounding the SPT protocluster has yielded an additional 15 (to the original 14) protocluster members, potentially doubling prior halo mass estimates \citep{hill2020}. This demonstrates that, until a thorough study on the larger scale of the structure is carried out, the true observed mass of the DRC (and other high-z protoclusters like it) will remain unknown. 

Given the high mass that appears in place for the DRC already at $z=4$, we consider next how the DRC may evolve compared to massive clusters seen locally today. Based on the evolutionary track for a Coma-like cluster derived in \citet[][grey line in Figure \ref{fig:dmh}]{chiang13}, we can generally estimate that the DRC will evolve to $\gtrsim10^{15}$\,M$_\odot$ by $z=0$. This is under the assumption that an overdensity such as the DRC (with $\delta_\mathrm{gal}>10,000\times$ the field density for massive galaxies, see Section \ref{sec:msevo}) traces one of, if not the, most massive halos in the large scale structure of the protocluster. We also derive a $z=0$ halo mass following \citet{lem14} by using the mean halo growth rate as a function of redshift and observed halo mass from the Millennium and Millenium-II simulations \citep{mcbride09,fakma10}. Using this method and the $z=4$ two different stellar-to-halo mass estimates outlined above, we derive a $z=0$ mass of M$_\mathrm{z=0} \approx 2-8\times10^{15}$\,M$_\odot$. This halo mass is extremely large, rivaling that of fully evolved galaxy clusters seen locally today \citep[e.g.][]{gitti04,gavazzi09}. Both halo mass evolution functions require a variety of assumptions of which we cannot constrain; e.g. the method used in \citet{lem14} was derived for a single halo, and the growth curves derived in \citet{chiang13} are highly dependent on the presumed volume of the observed galaxy overdensity. Considering these caveats, as well as a lack of additional constraints on the large scale structure of the DRC, and the uncertainties in stellar mass estimates, we state generally that the DRC is a massive cluster progenitor that will likely evolve $\gtrsim10^{15}$\,M$_\odot$ by $z=0$.

Overall, the list of factors that influence the future of this protocluster's growth is long, complex, and opaque (e.g. mergers, gas inflows, AGN, etc.). Still, with follow-up observations and simulation deep dives, we may be able to begin untangling the halo assembly past and future for massive cluster progenitors at $z>3$. Additional rest-frame UV/optical observations that map the extent of the DRC's large scale cluster would constrain the true mass distribution of the fated cluster. A deeper dive into simulations of massive cluster formation could shed light on halo mass configurations and distributions within protocluster galaxies, which can then be used to calibrate against standard abundance matching techniques for isolated halos. These efforts are left for future studies. 

\section{Summary and Conclusions} \label{sec:concl}
In this paper, we present a multiwavelength analysis on a $z=4.002$ SMG-rich, ultra-massive protocluster: the Distant Red Core. We combine new \textit{HST} and \textit{Spitzer} data with existing Gemini, \textit{Herschel}, and ALMA data to model spectral energy distributions for each respective ALMA object (Figure \ref{fig:seds}, except DRC-4), taking care to deblend low resolution \textit{Spitzer} IRAC data where needed (Section \ref{sec:tphot}). Stellar masses and SFRs are derived from SED-fitting with \textsc{cigale} (Section \ref{sec:SEDs}). Molecular gas mass estimates are derived using the observed-frame 2\,mm ALMA data (probing the Rayleigh-Jeans region of the dust continuum)  with the \citet{scoville} methodology. 

We confirm a population of massive (M$_\ast>10^{10}$\,M$_\odot$) galaxies in place when the Universe was only 1.5\,Gyr old. When comparing to field galaxies on SFR-M$_\ast$ plane (Figure \ref{fig:mainseq}), our results confirm that -- even at $z=4$ -- protocluster galaxies can be viewed as a high-mass (and possibly more bursty) extension of the star-forming main-sequence for coeval isolated field galaxies. Similarly, though several objects contain large gas reservoirs (M$_\mathrm{H_2} \gtrsim 10^{11}$\,M$_\odot$), all lie within the SFR-M$_\mathrm{H_2}$ main-sequence plane. When compared to $z=2-3$ protocluster and $z\sim4$ field counterparts, the DRC objects have similar gas mass fractions that follow the expected inverse f$_\mathrm{gas}-\mathrm{M_*}$ relationship. These systems also have short gas depletion timescales ($\sim260\pm180$\,Myr) on par with field SMGs which, in a closed box scenario, means that these objects will exhaust their gas supplies in time to become massive quiescent galaxies that dominate at cluster cores by $z\sim3$. 

Using multiple methods, we derive a total $z=4$ protocluster halo mass of $\sim10^{14}$\,M$_\odot$, and show that this value teeters on the edge of the most massive halo allowable/observable in the 600\,deg$^2$ \textit{H}-ATLAS survey volume (Figure \ref{fig:halos}). We estimate that the DRC will evolve to become an ultra-massive cluster with a total halo mass $>10^{15}$\,M$_\odot$ (possibly even $>10^{16}$\,M$_\odot$) at $z=0$ (Figure \ref{fig:dmh}). For both the $z=4$ and $z=0$ calculations, we argue that a more massive estimate may be appropriate based on the assumption that other significant galaxy populations within the protocluster's large scale structure are not included in this analysis. Still, even if additional protocluster members are confirmed, more multi-wavelength studies of $z>3$ DSFG-rich protoclusters combined with studies on the evolution of mass distributions and the gas duty cycle in cluster formation simulations are necessary to fully appreciate and characterize complex systems such as the Distant Red Core.

\acknowledgments

ASL thanks Emiliano Merlin and Michael Cooper for helpful discussions on protocluster and galaxy evolution, and Amanda Pagul for guidance in using \textsc{tphot}. A.S.L. also thanks Charlie Long, Patrick Long, Erini Lambrides, and the Department of Physics and Astronomy at University of California, Irvine for support throughout this work. JLW acknowledges support from an STFC Ernest Rutherford Fellowship (ST/P004784/2). CMC thanks the National Science Foundation for support through grants AST-1714528 and AST-1814034, the University of Texas at Austin College of Natural Sciences, and the Research Corporation for Science Advancement for a 2019 Cottrell Scholar Award sponsored by IF/THEN, an initiative of Lyda Hill Philanthropies. H.D. acknowledges financial support from the Spanish Ministry of Science, Innovation and Universities (MICIU) under the 2014 Ramn y Cajal program RYC-2014-15686 and AYA2017-84061-P, the later one co-financed by FEDER (European Regional Development Funds).

This research is based on observations made with the NASA/ESA Hubble Space Telescope obtained from the Space Telescope Science Institute, which is operated by the Association of Universities for Research in Astronomy, Inc., under NASA contract NAS 5â26555. These observations are associated with program GO 15464. This work is based [in part] on observations made with the \textit{Spitzer Space Telescope}, which was operated by the Jet Propulsion Laboratory, California Institute of Technology under a contract with NASA. Support for this work was provided by NASA through an award issued by JPL/Caltech. This paper makes use of the following ALMA data: 2013.1.00449.S, 2013.A.00014.S, and 2013.1.00449.S. ALMA is a partnership of ESO (representing its member states), NSF (USA) and NINS (Japan), together with NRC (Canada), MOST and ASIAA (Taiwan), and KASI (Republic of Korea), in cooperation with the Republic of Chile. The Joint ALMA Observatory is operated by ESO, AUI/NRAO and NAOJ. The National Radio Astronomy Observatory is a facility of the National Science Foundation operated under cooperative agreement by Associated Universities, Inc. This research made use of Astropy,\footnote{http://www.astropy.org} a community-developed core Python package for Astronomy \citep{astropy:2013, astropy:2018}.

\pagebreak 

\bibliographystyle{yahapj}
\bibliography{references}

\appendix \label{sec:appendix}
\restartappendixnumbering
In Section \ref{sec:spire}, we find that the \textit{Herschel} SPIRE deblending software XID+ \citep{xid} produces significant (S/N$\gtrsim2$) SPIRE detections at the individual level only when the four brightest 2\,mm sources are used as priors in the fit. Any larger combination of sources does not yield significant detections, indicating that the majority of the SPIRE flux can be contributed to these sources. If we were to be more conservative and instead use the SPIRE flux density estimates from the 11-object fit, the estimates for DRC objects 1-3 would be smaller and produce some different results (see Appendix Figure \ref{fig:1-4seds} and Table \ref{table:1-4obs}).

Overall, the infrared luminosities, star formation rates, stellar masses, and dust temperatures (all derived from SED-fitting) are smaller, but not significantly enough to change the conclusions drawn on DRC versus field galaxies. All three objects would still maintain ULIRG status (L$_\mathrm{IR} > 10^{12}$\,L$_\odot$), but with much cooler dust temperatures ($\langle\mathrm{T_d}\rangle = 21$\,K). DRC-1 would still remain relatively massive and bursty, but DRC-3 would become a significantly smaller, less star-forming galaxy, reducing by an order of magnitude in each respective feature. All three objects would remain on the SFR-M$_*$ main-sequence (though lower in stellar mass), as well as the SFR-M$_\mathrm{H_2}$ main-sequence. 

The reduction in combined stellar mass for these objects
has the largest impact on the total halo mass estimates for the cluster. Using the values reported in Appendix Table \ref{table:1-4obs}, cluster halo mass estimates, derived as either a single massive stellar halo or as many individual halos summed, are closer to the low end of those predicted in Section \ref{sec:dmh} and Figure \ref{fig:halos} at $\sim4\times10^{13}$\,M$_\odot$. This still results in a massive predicted $z=0$ halo size of $\sim10^{15}$\,M$_\odot$.

\begin{figure}[h]
\begin{center}
\includegraphics[trim=0.3cm 0.4cm 0cm 0cm, scale=1.75]{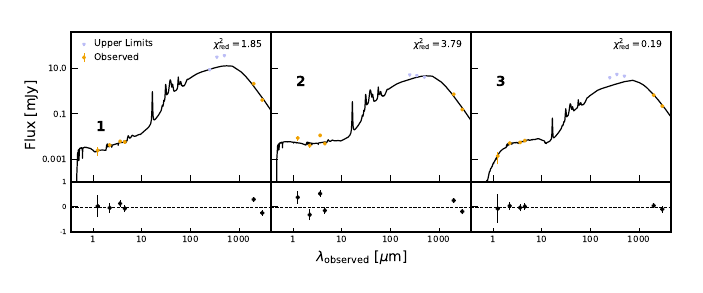}
\end{center}
\caption{Best fit spectral energy distributions from \textsc{cigale} using deblended SPIRE fluxes for \textit{all} DRC components. Symbols and colors are the same as in Figure \ref{fig:seds}. See Section \ref{sec:SEDs} for general SED fitting details.
\footnotesize 
}
\label{fig:1-4seds}
\end{figure}

\begin{deluxetable}{lcccccccccccccc}[h]
    \centering
    \tablecaption{Estimated flux densities for DRC 1-3 if the SPIRE deblended fit included all 11 DRC components, and galaxy properties from the resulting \textsc{cigale} SED fits. Measurements listed without uncertainties are used as upper limits. See Section \ref{sec:phot} and \ref{sec:spire} for more details on SPIRE photometry, and Section \ref{sec:SEDs} for general SED fitting details.}
    \tablehead{
    \colhead{ID} &
    \colhead{S$\mathrm{_{250\,\mu m}}$} & \colhead{S$\mathrm{_{350\,\mu m}}$} & \colhead{S$\mathrm{_{500\,\mu m}}$} & 
    \colhead{log(L$\mathrm{_{IR}}$)} & 
    \colhead{SFR} & 
    \colhead{log(M$_{*}$)} & 
    \colhead{T$_\mathrm{d}$} & 
    \\
     & \colhead{[mJy]} & \colhead{[mJy]} & \colhead{[mJy]} 
     & \colhead{[$\times 10^{12} \mathrm{L_\odot}$]} & \colhead{[M$_\odot$/yr]} & \colhead{[$\times 10^{10}$\,M$_\odot$]} & \colhead{[K]} 
     \\
     }
    \startdata
    DRC-1 & $<8.60$ & $<30.58$ & $<35.50$ & 12\,$\pm\,5$ & 1062\,$\pm\,890$ & 11\,$\pm\,8$ & 21 \\
    DRC-2 & $<5.07$ & $<4.72$ & $<4.03$ & 5\,$\pm\,1$ & 571\,$\pm\,320$ & 4\,$\pm\,2$ & 21 \\ 
    DRC-3 & $<3.78$ & $<5.28$ & $<4.47$ & 3\,$\pm\,1$ & 349\,$\pm\,233$ & 4\,$\pm\,4$ & 20\\
    \enddata
\end{deluxetable}\label{table:1-4obs}

\end{document}